\documentclass[journal]{IEEEtran}
\usepackage{cite}

\ifCLASSINFOpdf
  \usepackage[pdftex]{graphicx}
\else
\fi
\usepackage{amsmath}
\usepackage{amssymb}
\usepackage{bm}
\usepackage{flushend}
\usepackage{mathtools}

\usepackage{algcompatible}
\usepackage{algorithm}
\usepackage{algpseudocode}
\algrenewcommand\algorithmicindent{0em}
\makeatletter
\algnewcommand{\LineComment}[1]{\Statex \hskip\ALG@thistlm #1}

\makeatother
\usepackage{setspace}
\let\Algorithm\algorithm
\renewcommand\algorithm[1][]{\Algorithm[#1]\setstretch{1}}

\usepackage{color}
\newtheorem{Definition}{Definition}

\newtheorem{Lemma}{Lemma}
\newcommand{\Es}{E_{\mathrm{s}}}
\newcommand{\relmiddle}[1]{\mathrel{}\middle#1\mathrel{}}

\usepackage{tikz}

\newcommand\submittedtext{%
  \footnotesize This work has been submitted to the IEEE for possible publication. Copyright may be transferred without notice, after which this version may no longer be accessible.}

\newcommand\submittednotice{%
\begin{tikzpicture}[remember picture,overlay]
\node[anchor=south,yshift=10pt] at (current page.south) {\fbox{\parbox{\dimexpr0.65\textwidth-\fboxsep-\fboxrule\relax}{\submittedtext}}};
\end{tikzpicture}%
}

\ifCLASSOPTIONcompsoc
  \usepackage[caption=false,font=normalsize,labelfont=sf,textfont=sf]{subfig}
\else
  \usepackage[caption=false,font=footnotesize]{subfig}
\fi
\usepackage[acronym,shortcuts]{glossaries}
\usepackage{tcolorbox}
\definecolor{mycolor}{RGB}{242,242,242}

\newacronym{5G}{5G}{fifth generation}
\newacronym{6G}{6G}{sixth generation}
\newacronym{ADR}{ADR}{antenna decentralized rate}
\newacronym{AER}{AER}{activity error rate}
\newacronym{AP}{AP}{access point}
\newacronym{ADC}{ADC}{analog-to-digital converter}
\newacronym{ABP}{ABP}{approximate belief propagation}
\newacronym{ADD}{ADD}{annealed discrete denoiser}
\newacronym{AMP}{AMP}{approximate message passing}
\newacronym{AUD}{AUD}{active user detection}
\newacronym{AWGN}{AWGN}{additive white Gaussian noise}
\newacronym{BMMSE}{BMMSE}{Bussgang minimum mean square error}
\newacronym{BC}{BC}{belief combining}
\newacronym{BER}{BER}{bit error rate}
\newacronym{BiGAMP}{BiGAMP}{bilinear generalized approximate message passing}
\newacronym{BIP}{BIP}{bilinear inference problem}
\newacronym{GAMP}{GAMP}{generalized AMP}
\newacronym{GF}{GF}{grant-free}
\newacronym{BiGaBP}{BiGaBP}{bilinear Gaussian belief propagation}
\newacronym{BP}{BP}{belief propagation}
\newacronym{BS}{BS}{base station}
\newacronym{CAP}{CAP}{central AP}
\newacronym{CCU}{CCU}{central computing unit}
\newacronym{CDF}{CDF}{cumulative distribution function}
\newacronym{CLT}{CLT}{central limit theorem}
\newacronym{CPU}{CPU}{central processing unit}
\newacronym{CE}{CE}{channel estimation}
\newacronym{CF-mMIMO}{CF-mMIMO}{cell-free massive MIMO}
\newacronym{CSI}{CSI}{channel state information}
\newacronym{CSIDCO}{CSIDCO}{complex SIDCO}
\newacronym{DCC}{DCC}{dynamic cooperation clustering}
\newacronym{DFT}{DFT}{discrete Fourier transform}
\newacronym{DoF}{DoF}{degrees of freedom}
\newacronym{DQ}{DQ}{De-quantization}
\newacronym{DL}{DL}{deep learning}
\newacronym{DU}{DU}{deep unfolding}
\newacronym{eMBB}{eMBB}{enhanced mobile broadband}
\newacronym{ECF}{ECF}{estimate-compress-forward}
\newacronym{EP}{EP}{expectation propagation}
\newacronym{EPA}{EPA}{approximate EP}
\newacronym{FA}{FA}{false alarm}
\newacronym{FN}{FN}{factor node}
\newacronym{FG}{FG}{factor graph}
\newacronym{FTN}{FTN}{Faster-than-Nyquist}
\newacronym{GaBP}{GaBP}{Gaussian belief propagation}
\newacronym{IC}{IC}{interference cancellation}
\newacronym{IDD}{IDD}{iterative detection and decoding}
\newacronym{i.i.d.}{i.i.d.}{independent and identically distributed}
\newacronym{JACDE}{JACDE}{joint activity, channel and data estimation}
\newacronym{JACE}{JACE}{joint activity and channel estimation}
\newacronym{JCDE}{JCDE}{joint channel and data estimation}
\newacronym{KLD}{KLD}{Kullback-Leibler divergence}
\newacronym{LE}{LE}{linear estimator}
\newacronym{LSA}{LSA}{latent semantic analysis}
\newacronym{LMMSE}{LMMSE}{linear MMSE}
\newacronym{MAC}{MAC}{multiple-access channel}
\newacronym{MAP}{MAP}{maximum \textit{a posteriori} probability}
\newacronym{MPDQ}{MPDQ}{message passing DQ}
\newacronym{MD}{MD}{miss-detection}
\newacronym{MF}{MF}{matched filter}
\newacronym{MFB}{MFB}{matched filter bound}
\newacronym{MF-EP}{MF-EP}{matched filter EP}
\newacronym{MM}{MM}{moment matching}
\newacronym{MNS}{MNS}{minimum norm solution}
\newacronym{MIMO}{MIMO}{multi-input multi-output}
\newacronym{MU-MIMO}{MU-MIMO}{multi-user MIMO}
\newacronym{MU}{MU}{multi-user}
\newacronym{mMIMO}{mMIMO}{Massive multiple-input multiple-output}
\newacronym{ML}{ML}{maximum likelihood}
\newacronym{MMSE}{MMSE}{minimum mean square error}
\newacronym{mMTC}{mMTC}{massive machine type communications}
\newacronym{MMV-AMP}{MMV-AMP}{multiple measurement vector approximate message passing}
\newacronym{MSE}{MSE}{mean square error}
\newacronym{MUD}{MUD}{multi-user detection}
\newacronym{NLE}{NLE}{nonlinear estimator}
\newacronym{NR}{NR}{new radio}
\newacronym{NMSE}{NMSE}{normalized mean square error}
\newacronym{OAMP}{OAMP}{orthogonal AMP}
\newacronym{OFDM}{OFDM}{orthogonal frequency-division multiplexing}
\newacronym{PDA}{PDA}{probabilistic data association}
\newacronym{PDF}{PDF}{probability density function}
\newacronym{PMF}{PMF}{probability mass function}
\newacronym{PPP}{PPP}{Poisson point process}
\newacronym{PSK}{PSK}{phase-shift keying}
\newacronym{QP}{QP}{quadratic program}
\newacronym{QPSK}{QPSK}{quadrature PSK}
\newacronym{QAM}{QAM}{quadrature amplitude modulation}
\newacronym{SIDCO}{SIDCO}{sequential iterative decorrelation via convex optimization}
\newacronym{SD}{SD}{sphere decoding}
\newacronym{SE}{SE}{state evolution}
\newacronym{SGA}{SGA}{scalar Gaussian approximation}
\newacronym{SIC}{SIC}{soft interference cancellation}
\newacronym{SID}{SID}{self-iterative detection}
\newacronym{SNR}{SNR}{signal-to-noise ratio}
\newacronym{Soft IC}{Soft IC}{soft interference cancellation}
\newacronym{SotA}{SotA}{state-of-the-art}
\newacronym{SVD}{SVD}{singular value decomposition}
\newacronym{SPA}{SPA}{sum-product algorithm}
\newacronym{TX}{TX}{transmit}
\newacronym{RX}{RX}{receive}
\newacronym{UE}{UE}{user equipment}
\newacronym{ULA}{ULA}{uniform linear array}
\newacronym{URA}{URA}{unsourced random access}
\newacronym{URLLC}{URLLC}{ultra reliable low latency communications}
\newacronym{VAMP}{VAMP}{vector AMP}
\newacronym{VGA}{VGA}{vector Gaussian approximation}
\newacronym{VN}{VN}{variable node}
\newacronym{w.r.t.}{w.r.t.}{with respect to}
\newacronym{ZF}{ZF}{zero-forcing}
\newacronym{flops}{flops}{floating point operations}
\newacronym{CS}{CS}{compressed sensing}
\newacronym{MP}{MP}{message passing}
\newacronym{MPA}{MPA}{message passing algorithm}
\newacronym{DNN}{DNN}{deep neural network}
\newacronym{ASB}{ASB}{adaptively scaled belief}
\newacronym{LLR}{LLR}{log-likelihood ratio}
\newacronym{BAd-VAMP}{BAd-VAMP}{bilinear adaptive vector AMP}
\newacronym{LIP}{LIP}{linear inference problem}
\newacronym{AoA}{AoA}{angle of arrival}
\newacronym{LS}{LS}{least square}
\newacronym{mmWave}{mmWave}{millimeter-wave}

\renewcommand{\baselinestretch}{0.91}

\usepackage{comment}

\begin{document}
%
%
\title{
Discrete-Valued Signal Estimation \\via Low-Complexity Message Passing Algorithm
\\ for Highly Correlated Measurements}
%
%
%

\author{
Tomoharu~Furudoi,~\IEEEmembership{Graduate Student Member,~IEEE},
~Takumi~Takahashi,~\IEEEmembership{Member,~IEEE},\\
~Shinsuke~Ibi,~\IEEEmembership{Senior Member,~IEEE},
and~Hideki~Ochiai,~\IEEEmembership{Fellow,~IEEE},
\thanks{

T. Furudoi, T. Takahashi and H. Ochiai are with Graduate School of Engineering, Osaka University 2-1 Yamada-oka, Suita, 565-0871, Japan (e-mail: furutomo@wcs.comm.eng.osaka-u.ac.jp, \{takahashi, ochiai\}@comm.eng.osaka-u.ac.jp).

S. Ibi is with Faculty of Science and Engineering, Doshisha University 1-3 Tataramiyakodani, Kyotanabe, 610-0394, Japan (e-mail: sibi@mail.doshisha.ac.jp).}
\vspace{-7mm}
}

%
%

\markboth{Journal of \LaTeX\ Class Files,~Vol.~, No.~, November~2024}%
{T.~Takahashi \MakeLowercase{\textit{et al.}}: Bare Demo of IEEEtran.cls for IEEE Journals}
%



\maketitle

\submittednotice

\vspace{-7mm}
\begin{abstract}
This paper considers a discrete-valued signal estimation scheme based on a low-complexity Bayesian optimal \ac{MPA} for solving massive linear inverse problems under highly correlated measurements.
\Ac{GaBP} can be derived by applying the \ac{CLT}-based Gaussian approximation to the \ac{SPA} operating on a dense \ac{FG}, while \ac{MF}-\ac{EP} can be obtained based on the \ac{EP} framework tailored for the same \ac{FG}.
Generalized approximate message passing (GAMP) can be found by applying a rigorous approximation technique for both of them in the large-system limit, and these three \acp{MPA} perform signal detection using \ac{MF} by assuming large-scale uncorrelated observations.
However, each of them has a different inherent self-noise suppression mechanism, which makes a significant difference in the robustness against the correlation of the observations when we apply an \textit{\ac{ADD}} that adaptively controls its nonlinearity with the inverse temperature parameter corresponding to the number of iterations.
In this paper, we unravel the mechanism of this interesting phenomenon, and further demonstrate the practical applicability of the low-complexity Bayesian optimal \ac{MPA} with \ac{ADD} under highly correlated measurements.

\end{abstract}


\glsresetall

\IEEEpeerreviewmaketitle

\vspace{-2mm}
\section{Introduction}

Consider a discrete-valued signal estimation from a noisy linear measurement expressed as
\begin{equation}
    \label{equ: linear_measurement}
    \bm{y} = \bm{A}\bm{x} + \bm{w},
\end{equation}
where $\bm{x} \triangleq \left[x_1, x_2, \ldots , x_M \right]^{\mathsf{T}} \in \mathcal{X}^{M \times 1}$ denotes an unknown discrete-valued signal vector, each element of which is selected independently from a finite countable set $\mathcal{X}$,
$ \bm{A} \in \mathbb{C}^{N \times M} $ denotes a known measurement matrix, and 
$\bm{w}  
$ denotes a complex
circularly symmetric\footnote{
A random variable $ X $ is  \textit{circularly symmetric} when $ X $ and $ Xe^{\mathrm{j}\theta} $ follow the same distribution for all $ \theta \in [0, 2\pi ) $~\cite{Picinbono1994}.
}
\ac{i.i.d.} \ac{AWGN} vector, each element of which has zero mean and
variance $\sigma_{\mathrm{w}}^2$.
Many estimation problems in various engineering fields, such as physical layer signal processing in wireless communications~\cite{Mazo1975, Verdu1998, Mohammed2009, Chockalingam2014, Yang2015, Hayakawa2018, Yang2024} and digital image processing~\cite{Duarte2011, Bioglio2014, Sarangi2022}, can be formulated as discrete-valued signal estimation problems based on \eqref{equ: linear_measurement}.
Our goal in this paper is to estimate the unknown discrete-valued vector $\bm{x}$ based on perfect knowledge of $ \bm{y}, \, \bm{A}, \, \sigma_{\mathrm{w}}^2$, and a prior probability distribution of $\bm{x}$.

As is well known, the optimal discrete-valued signal estimation scheme is based on the \ac{MAP} or \ac{MMSE} criterion, which are, however, practically infeasible in a large system since these methods are essentially equivalent to an exhaustive search over the entire space of $\mathcal{X}^{M \times 1}$ and lead to an exponentially growing computational complexity, \textit{i.e.}, $\mathcal{O} \left( |\mathcal{X}|^M \right) $.
%
In contrast, linear filters, such as \ac{MF} and \ac{LMMSE} filter, are often employed as naive low-complexity detection schemes, but they can achieve favorable performance only in sufficiently \textit{overdetermined} (\textit{i.e.}, $ N \gg M $) systems and when $N$ and $M$ are of the same order, their performance inevitably degrades~\cite{Chockalingam2014}. 

In order to achieve large-scale linear inference (\textit{i.e.}, $M,N\gg 1$) with low-complexity and high-accuracy, various \acp{MPA} have been investigated~\cite{Kabashima2003, Takahashi2019_tcom, Donoho2009, Rangan2011, Cespedes2014, Meng2015, Meng2018, Rangan2019, Takeuchi2020, Liu2022, Takeuchi2021}.
Among these, \ac{AMP}~\cite{Donoho2009} and its extension to general linear observations, \textit{i.e.}, \ac{GAMP}~\cite{Rangan2011},
have attracted considerable attention from both theoretical and practical perspectives.
The remarkable advantage of \ac{AMP} is its ability to asymptotically converge to the Bayesian optimal (\textit{i.e.}, \ac{MMSE}) solution in the large-system limit for arbitrary prior distributions with minimal computational complexity of $\mathcal{O}(MN)$ per iteration, provided that $\bm{A}$ is composed of \ac{i.i.d.} Gaussian random variables with mean zero and its \ac{SE} has a unique fixed point~\cite{Bayati2011, Bayati2015, Takeuchi2019}.
The most well-known derivation of \ac{AMP} is a method for approximating the \ac{SPA} operating on a dense \ac{FG} in the large-system limit\footnote{
The idealized system assumption, where the input and output dimensions,
 $M$ and $N$, respectively, are infinity for a given compression ratio $ \xi \triangleq N/M $.
}~\cite{Bayati2011}.
According to \cite{Tamaki2022}\footnote{This conference paper is an earlier version of this paper,  which was presented at the IEEE ICC 2022.}, the derivation process can be divided into 1) a Gaussian approximation of the message based on the \ac{CLT} and 2) a large-system limit approximation of the message moments, and the \ac{MPA} obtained by completing step 1) is equivalent to the very well-known naive \ac{BP} algorithm called \ac{ABP}~\cite{Kabashima2003} or \ac{GaBP}~\cite{Takahashi2019_tcom}.
In other words, \ac{GaBP} is found as an intermediate step in deriving \ac{AMP} from \ac{SPA}, and \ac{AMP} can be systematically derived in the large-system limit of \ac{GaBP} under the ideal statistical assumption.

\Ac{EP}~\cite{Minka2001, Opper2005, Bishop2006} is another well-known framework that derives \ac{AMP} differently from \ac{BP}.
\ac{EP} is one of the approximate inference frameworks first proposed in~\cite{Minka2001}. 
It approximates the true posterior distribution of an unknown signal with a tractable distribution belonging to the exponential distribution family by minimizing their \ac{KLD}.
A plethora of \acp{MPA} based on \ac{EP} have been proposed~\cite{Cespedes2014, Meng2015, Meng2018, Takeuchi2020, Santos2020, Zeng2024}, including \ac{OAMP}~\cite{Ma2017} and \ac{VAMP}~\cite{Rangan2019}, and rigorous theoretical analyses of their dynamics have been provided~\cite{kabashima2014, Cakmak2019}.
According to these results, the \ac{EP} algorithm operating on a dense \ac{FG} can also be interpreted as approximated \ac{BP} with the minimal complexity order, and their message update rules are derived by projecting the posterior distribution of the unknown signal onto the Gaussian distribution based on \textit{\ac{MM}} technique.
Henceforth, we will refer to this algorithm as \ac{MF-EP} so as to distinguish it from the \ac{LMMSE}-based \ac{EP} algorithm where the latter guarantees Bayesian optimality for unitarily invariant observations\cite{Rangan2019, Takeuchi2020}.
It is worth noting here that, as expected, \ac{AMP} can be rigorously derived by applying a large-system limit approximation to \ac{MF-EP}~\cite{Meng2015}.
Under the ideal statistical assumption of \ac{i.i.d.} zero-mean Gaussian measurements, all Bayesian optimal \acp{MPA} with the minimal complexity order of $\mathcal{O}(MN)$ converge to a
unique form of (G)\ac{AMP} in the large-system limit\cite{Bayati2011,Bayati2015, Takeuchi2019}.

Bayesian optimal \ac{MPA} always has a mechanism to suppress (decouple) the self-noise feedback due to its iterative structure, which enables convergence to the optimal solution through the exchange of \textit{extrinsic} information across iterations.
The key difference among \ac{GaBP}, \ac{MF-EP}, and \ac{GAMP} lies in the method of generating this extrinsic information.
\ac{GaBP} inherits the algorithmic structure of \ac{SPA}, where each of the nodes comprising the \ac{FG} generates extrinsic information by combining the likelihood information excluding feedback components from its own nodes~\cite{Kabashima2003,Takahashi2019_tcom}.
In \ac{MF-EP}, in contrast to \ac{GaBP}, each node fully combines all propagated likelihood information, including its own, and after projecting it into a tractable distribution, the extrinsic information is generated by a division operation on the resulting message domain~\cite{Meng2015}.
In the large-system limit condition, these two mechanisms for suppressing self-noise feedback converge on a mechanism called \textit{Onsager correction}, which predicts and decouples the effect of self-noise feedback across iterations in \ac{GAMP}~\cite{Donoho2009,Rangan2011}.

As would be expected from their Bayesian optimality, these three algorithms exhibit similar behavior and achieve comparable estimation accuracy under conditions of (nearly ideal) large-scale uncorrelated observations. 
Nevertheless, many practical systems do not preserve such ideal conditions, and thus their behaviors become analytically less tractable. In fact, due to their unique self-noise suppression mechanisms, these \acp{MPA} exhibit substantially different levels of robustness against correlated observations as we demonstrate in this paper.
Based on the study on this intriguing phenomenon, we aim to propose an advanced and effective strategy for achieving accurate discrete-valued signal estimation with minimal complexity even under highly correlated measurements, challenging the conventional understanding that low-complexity \acp{MPA} only work under the idealized and limited condition. 

To assist the generation of extrinsic information under non-ideal conditions, the key technique used in this work is \textit{belief scaling}~\cite{Takahashi2019_tcom}.
The reason why the performance of these \acp{MPA} deteriorates significantly under highly correlated measurements is that when the estimated (Gaussian) distribution of the propagated messages (\textit{i.e.}, \textit{beliefs}) cannot sufficiently describe the actual stochastic behavior of the effective noise, the accuracy of the conditional expectation (\textit{i.e.}, \textit{soft replicas}) computed by a \textit{denoiser} deteriorates due to model errors~\cite{Takahashi2019_ieice, Caltagirone2014, Rangan2019conv}.
In particular, for discrete-valued signal estimation, the denoiser has nonlinearity due to the discrete constraints of the unknown signal, but if the shape (\textit{softness} of the denoiser curve) deviates from an appropriate value due to the model error of the beliefs, severe error propagation occurs in the early iterations, considerably hindering the exchange of proper extrinsic information and 
resulting in low estimation accuracy.
As a highly effective solution to mitigate such potential issues, belief scaling has been proposed in the context of \ac{MIMO} detections~\cite{Takahashi2019_tcom}, which controls the convergence speed by adaptively adjusting the nonlinearity of the denoiser using parameters according to belief variance (\textit{i.e.}, instantaneous reliability) and to the predetermined number of iterations.
This methodology has been shown to be highly effective for improving poor convergence properties caused by finite-sized practical system configurations~\cite{Ito2024,Takahashi2023,Takahashi2022,Iimori2021}, and/or a mild correlation in observations~\cite{Shirase2020,Shirase2021}, and is generalized in this paper as an \textit{\ac{ADD}} that is independent of the discrete prior of the unknown signal and/or the employed \ac{MPA}.

Based on the above, in this paper, we consider the comparative study of the performance differences among \ac{GaBP}, \ac{MF-EP}, and \ac{GAMP} adopting \ac{ADD} in the estimation of discrete-valued signal under highly correlated measurements, both theoretically and numerically.
Surprisingly, numerical simulations show that both \ac{MF-EP} and \ac{GAMP} with \ac{ADD} significantly improve the performance in highly correlated measurements, comparable to or even superior to \ac{LMMSE}-\ac{EP}, which requires iterative \ac{LMMSE} filtering.
It is worth noting here that such a dramatic improvement is not seen in \ac{GaBP}, which means that these three \acp{MPA}, whose structures asymptotically converge under the idealized large-system limit condition, exhibit completely different robustness against measurement correlations when using \ac{ADD}. 
This astonishing performance improvement observed in \ac{MF-EP} and \ac{GAMP} is due to their potential ability to exchange accurate extrinsic information even under correlated observations with the aid of \ac{ADD}.
To unravel the mechanism of this interesting phenomenon, we will formulate a theoretical hypothesis based on the algorithmic structure, and then verify it by introducing a matrix that visualizes the statistical correlation structure between the propagated beliefs, and by studying its dynamics across iterations, as well as the actual behavior of beliefs.

The contributions of this paper are summarized as follows:
\begin{itemize}
\item Assuming that each element of $\bm{A}$ follows a circularly symmetric distribution, we describe the detailed process of deriving \ac{GAMP} from \ac{GaBP} for an arbitrary (not limited to discrete-valued) unknown vector $\bm{x}$.
The derivation process presented here is an extension of the derivation in~\cite{Tamaki2022}, which assumes a real-valued system, to a complex-valued one using the Wirtinger derivative and Taylor expansion. 
A similar derivation approach is presented in~\cite{Donoho2010} for the real-valued system, and in~\cite{ Ito2024,Fujitsuka2020} for bilinear inference, but to the best of our knowledge, there is no direct derivation from \ac{GaBP} to \ac{GAMP} in general complex-valued systems, which would provide a theoretical basis for elucidating the relationship among \ac{GaBP}, \ac{MF-EP}, and \ac{GAMP}.
\item By interpreting the scaling parameter as an inverse temperature parameter corresponding to the number of iterations, the belief scaling method proposed in the context of \ac{GaBP}-based \ac{MIMO} detection \cite{Takahashi2019_tcom} can be generalized as the \ac{ADD} that is independent of the discrete prior of the unknown signal and/or the employed \ac{MPA}.
Furthermore, we compare the discrete-valued signal estimation performance of \ac{GaBP}, \ac{MF}-\ac{EP}, and \ac{GAMP} using \ac{ADD} in a simulation of correlated massive \ac{MIMO} detection, and the results reveal that 1) the performance of the three algorithms is quite different and 2) \ac{MF}-\ac{EP} and \ac{GAMP} achieve better performance than \ac{LMMSE}-\ac{EP} even in highly correlated measurements. 
\item In order to elucidate the differences in signal detection performance and the dramatic performance improvements observed in the above simulations, we analyze the behavior of each algorithm in terms of the self-noise suppression mechanism and perform numerical studies to verify the hypothesis derived from this analysis.
The numerical analysis of the algorithm dynamics reveals that the timing of self-noise suppression in the iterative process, \textit{i.e.}, the relative position \ac{w.r.t.} the denoiser, significantly changes the correlation structure between beliefs, allowing to describe the dramatic performance improvement in \ac{MF}-\ac{EP} and \ac{GAMP} due to \ac{ADD}.
\end{itemize}

\textit{Notation}:
Sets of real and complex numbers are denoted by $\mathbb{R}$ and $\mathbb{C}$.
Vectors and matrices are denoted in lower- and upper-case bold-face fonts, respectively.
The conjugate and transpose
are denoted by $(\cdot)^*$ and $(\cdot)^{\mathsf{T}}$, 
respectively.
The real and imaginary parts of a complex quantity are denoted by $ \Re[\cdot]$ and $\Im[\cdot]$, and the imaginary unit by $\mathrm{j} \triangleq \sqrt{-1}$.
The $K \times K$ identity matrix
is denoted by
$ \bm{I}_K$. 
%
For any countable finite set $\mathcal{A}$, the number of elements in $\mathcal{A}$ is denoted by $|\mathcal{A}|$.
The $i$-th element of a vector $\bm{a}$ and the $(i, j)$-th element of a matrix $\bm{A}$ are denoted by $[\bm{a}]_i$ and $[\bm{A}]_{i, j}$, respectively.
The complex circularly symmetric Gaussian distribution with a mean vector $\bm{\mu}$ and a covariance matrix $\bm{\varLambda}$ is denoted by $\mathcal{CN}\left(\bm{\mu}, \bm{\varLambda}\right)$.
%
The notation $a \sim \mathcal{P}$ indicates a random variable $a$ follows a probability distribution $ \mathcal{P} $.
The \ac{PMF}, \ac{PDF}, and expectation \ac{w.r.t.} random variable $a$ are denoted by $P_{\mathsf{a}}[\cdot],  \, p_{\mathsf{a}}(\cdot),$ and
$\mathbb{E}_{\mathsf{a}}[\cdot]$, respectively.
In addition, the conditional \ac{PDF} and expectation \ac{w.r.t.} $a$ given the realized value $b'$ of random variable $b$ are denoted by 
$ p_{\mathsf{a} \mid \mathsf{b}}(\cdot \mid b') $ and $ \mathbb{E}_{\mathsf{a} \mid \mathsf{b} = b'}[\cdot] $, respectively.
For brevity, we use the 
notation 
$\sum_{i \neq j}^{I} a_{i}
\triangleq
\sum_{i=1}^{I} a_i - a_j$, 
$ \prod_{i \neq j}^{I} a_{i}
\triangleq
\left(\prod_{i=1}^{I} a_i\right) / a_j$, and 
$ \int f(\bm{a}) \,
\mathrm{d}\bm{a}_{\backslash j}
\triangleq
\int \cdots \int f(\bm{a}) \, \prod_{i \neq j}^{I}\mathrm{d}a_i$
for a multiple integral of any function $f: \mathbb{C}^{I \times 1} \to \mathbb{R}$ \ac{w.r.t.} all the elements in vector $\bm{a} = [a_1, a_2, \ldots , a_I]^{\mathsf{T}} \in \mathbb{C}^{I \times 1}$ except for $a_j$.
Finally, the notation $\mathcal{O}(\cdot)$ denotes the complexity order unless otherwise specified.

\section{Preliminaries}

\subsection{Signal Model}
Consider the complex-valued signal model in \eqref{equ: linear_measurement}, where the \ac{PDF} of the discrete-valued signal $x_m, \, \forall m \in \left\{ 1, 2, \cdots , M \right\}$, can be expressed as
\begin{equation}
    \label{equ: p_xm_w_disc_prior}
    p_{\mathsf{x}_m} (x)
    =
    \sum_{\chi \in \mathcal{X}}
    P_{\mathsf{x}_m}[\chi] 
    \cdot
    \delta (\chi - x),
\end{equation}
with $ \delta (\cdot) $ denoting Dirac measure.
Without loss of generality, we assume 
$ \mathbb{E}_{\bm{\mathsf{x}}} [\bm{x}] = \bm{0} $ and
$ \mathbb{E}_{\bm{\mathsf{x}}} \left[\bm{x} \bm{x}^{\mathsf{H}} \right] = \sigma_{\mathrm{x}}^2 \bm{I}_M $.
For later convenience, we define 
$ y_n \triangleq [\bm{y}]_n, \, w_n \triangleq [\bm{w}]_n,$ and $a_{n, m} \triangleq [\bm{A}]_{n, m}$ for $m \in \left\{1, 2, \cdots , M \right\}$ and $ n \in \left\{1,2 \cdots , N \right\} $.

\subsection{Bayesian Optimal Denoiser for AWGN Channels}

Bayesian optimal \acp{MPA} usually consist of two modules: one is a \ac{LE} that performs signal separation, and the other is a \ac{NLE} that computes the conditional expectation (\textit{i.e.}, soft replica) according to the prior distribution of the unknown signal.
In \ac{NLE}, the \ac{MMSE} estimate is computed by (approximately) modeling each of the signals separated by \ac{LE} as a virtual \ac{AWGN} channel output. Hence, we will introduce the mathematical representation of \ac{NLE} for \ac{AWGN} channels, \textit{i.e.}, the Bayesian optimal denoiser, and its vital property.

\begin{Definition}[Bayesian optimal denoiser]
\label{def: bayes_optial_denoiser}
A function $\eta:  \mathbb{C} \to \mathbb{C} $ which generates the \ac{MMSE} estimate of $x$ 
from the \ac{AWGN}-corrupted signal 
\begin{equation}
    \label{equ: awgn}
    y = x + w,
    \quad
    w \sim \mathcal{CN}(0, v),
\end{equation}
is called \textit{Bayesian optimal denoiser}, which is defined by the conditional expectation as
\begin{equation}
        \label{equ: bayes_optimal_denoiser}
        \eta \left( y ; v \right)
        \triangleq
        \mathbb{E}_{\mathsf{x} \mid \mathsf{y} = y} [x]
        =
        \dfrac{
        \displaystyle
        \int_{\mathcal{I}} {x \cdot 
        p_{\mathsf{x}} (x) 
        e^{ - \frac{\left| y - x \right|^2 }{v}}
        \, \mathrm{d} x
        }}{
        \displaystyle
        \int_{\mathcal{I}} {
        p_{\mathsf{x}} (x') 
        e^{ - \frac{\left| y - x' \right|^2 }{v}}
        }
        \, \mathrm{d} x' },
    \end{equation}
\end{Definition}
where $\mathcal{I}$ is the support of the prior distribution $ p_{\mathsf{x}}(\cdot) $.
Here, we introduce the following lemma:
\begin{Lemma}
\label{lem: denoiser_identities}
For an arbitrary prior $ p_{\mathsf{x}}(\cdot) $,  the following identity holds \ac{w.r.t.}  $ \eta $ in \eqref{equ: bayes_optimal_denoiser}.
%
    \label{equ: denoiser_identity}
    \begin{eqnarray}
        \label{equ: denoiser_mse}
        v \cdot \frac{\partial \eta(y; v)}{\partial y}
        \!\!\!\!
        &=&
        \!\!\!\!
        \mathbb{E}_{\mathsf{x} \mid \mathsf{y} = y}
\left[ \left| x \right|^2 \right]
-
\left|
\mathbb{E}_{\mathsf{x} \mid \mathsf{y} = y}
\left[ x \right]
\right|^2,
    \end{eqnarray}
    %
    where Wirtinger derivative
    \begin{equation}
        \frac{\partial }{ \partial z}
        =
        \frac{1}{2 }
        \left(\frac{\partial }{ \partial z^\Re} - \mathrm{j}\frac{\partial }{ \partial z^\Im} \right)
        ,
        \,\,\,
        \frac{\partial }{ \partial z^*}
        =
        \frac{1}{ 2  }
        \left(\frac{\partial }{ \partial z^\Re} + \mathrm{j}\frac{\partial }{ \partial z^\Im} \right),
    \end{equation}
    for $ z = z^\Re + \mathrm{j} z^{\Im} \in \mathbb{C}$ is employed.
\end{Lemma}
\begin{IEEEproof}
By multiplying 
$ \int_{\mathcal{I}} {
p_{\mathsf{x}} (x) 
e^{ - \frac{\left| y - x \right|^2 }{v}}
}
\, \mathrm{d} x $
on both sides of \eqref{equ: bayes_optimal_denoiser} and then differentiating \ac{w.r.t.} $y$, we have
\begin{eqnarray}
&&
\!\!\!\!
\frac{\partial \eta(y; v)}{\partial y}
\cdot
\int_{\mathcal{I}} {
p_{\mathsf{x}} (x) 
e^{ - \frac{\left| y - x \right|^2 }{v}}}
\, \mathrm{d} x
\nonumber
\\
\!\!\!\!&&\!\!\!\!
\qquad \qquad
+
\,
\eta(y; v)
\cdot
\int_{\mathcal{I}} {
\frac{x^* - y^*}{v}
\cdot
p_{\mathsf{x}} (x) 
e^{ - \frac{\left| y - x \right|^2 }{v}}}
\, \mathrm{d} x
\nonumber
\\
\!\!\!\!&=&\!\!\!\!
\int_{\mathcal{I}} {
\frac{x \left( x^* - y^* \right)}{v}
\cdot
p_{\mathsf{x}} (x) 
e^{ - \frac{\left| y - x \right|^2 }{v}}}
\, \mathrm{d} x.
\end{eqnarray}
Multiplying
$v /  \int_{\mathcal{I}} {
p_{\mathsf{x}} (x) 
e^{ - \frac{\left| y - x \right|^2 }{v}}
}
\, \mathrm{d} x$ 
on both sides yields
\begin{eqnarray}
&&
\!\!\!\!\!\!\!\!
v \cdot \frac{\partial \eta(y; v)}{\partial y}
+
\eta (y; v) \left( \eta^*(y; v) - y^* \right)
\nonumber
\\
\!\!\!\!&&\!\!\!\!
\qquad \qquad \qquad 
= \mathbb{E}_{\mathsf{x} \mid \mathsf{y} = y}
\left[ |x|^2 \right]
-
\eta (y; v)y^*.
\end{eqnarray}
From the definition $ \eta(y; v) \triangleq  \mathbb{E}_{\mathsf{x} \mid \mathsf{y} = y} [x]$, we arrive at \eqref{equ: denoiser_mse}.
%
\end{IEEEproof}

From \eqref{equ: denoiser_mse}, we can compute the \ac{MSE} of the posterior estimate, \textit{i.e.}, $ \mathbb{E}_{\mathsf{x} \mid \mathsf{y} = y} [|x - \eta(y; v)|^2] $, via differentiation of $ \eta(y; v) $ regardless of prior $p_{\mathsf{x}}(\cdot)$. 
Thus, the following derivation in Sections \ref{chap: gabp_and_mfep} and \ref{chap: derivation_of_gamp} does not depend on the prior distribution for $\bm{x}$ in \eqref{equ: linear_measurement}.

\begin{figure*}[t]
\begin{center}
\subfloat[Belief from FN to VN]{
	\includegraphics[width=0.75\columnwidth,keepaspectratio=true]{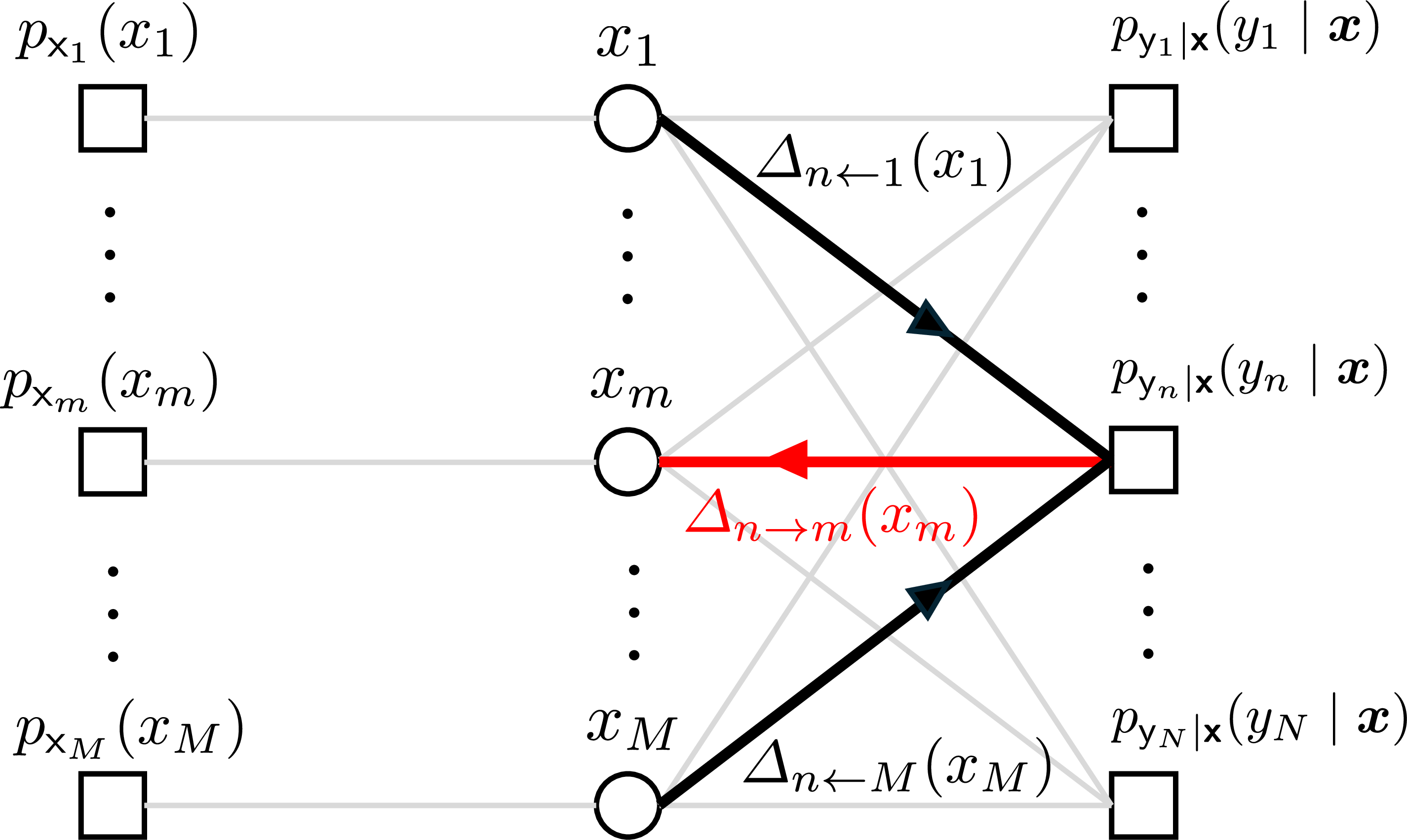}
	\label{fig: fg_spa_fn}
	}
 	\hspace{0mm}
	\subfloat[Belief from VN to FN]{
	\includegraphics[width=0.75\columnwidth,keepaspectratio=true]{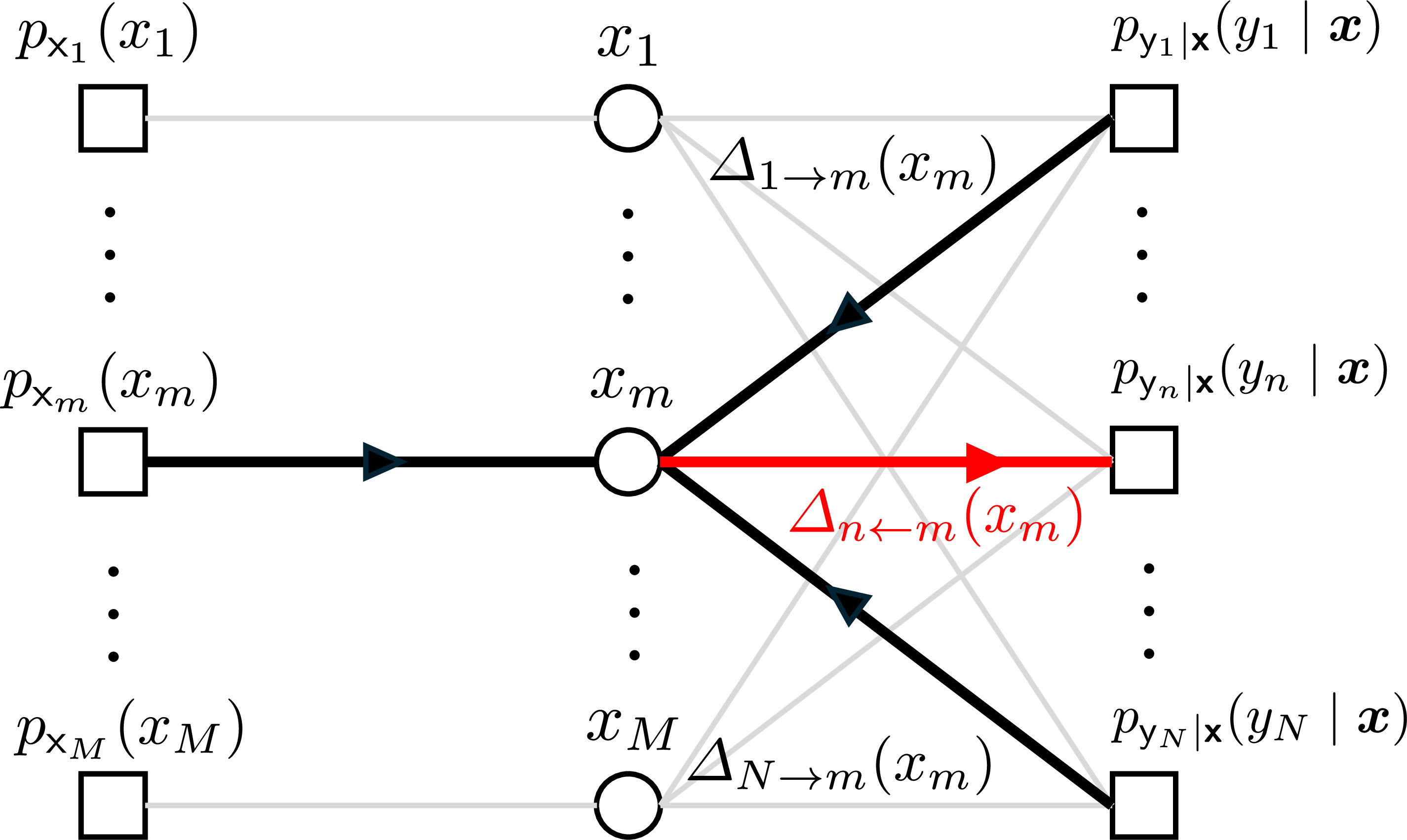}
	\label{fig: fg_spa_vn}
	}
	\caption{The illustration of update rules of \ac{SPA} on the \ac{FG}.}
	\label{fig: fg_spa}
	\vspace{-4mm}
\end{center}
\end{figure*}

\section{Derivation of GaBP and MF-EP}
\label{chap: gabp_and_mfep}

In this section, we derive two different Bayesian optimal \acp{MPA} referred to as \ac{GaBP} and \ac{MF}-\ac{EP}, and elucidate the differences in their structures from the perspective of extrinsic information generation through self-noise suppression.
\subsection{Derivation of GaBP from SPA}

Fig. \ref{fig: fg_spa} shows the \ac{FG} consisting of \acp{FN} and \acp{VN}, which correspond to the \acp{PDF} and the unknown signals, respectively.
The edges between nodes indicate dependencies according to $\bm{A}$.
\ac{GaBP} can be derived by applying \ac{CLT} to \ac{SPA} operating on the \ac{FG}.

First, denoting the belief propagated from the $n$-th \ac{FN} to the $m$-th \ac{VN} by $ \varDelta_{n \to m} (x_m) $ and one propagated in the opposite direction by $ \varDelta_{n \leftarrow m} (x_m) $, the update rule of \ac{SPA} can be expressed as
\begin{subequations}
    \label{equ: spa_messages}
\begin{eqnarray}
    \!\!\!\!\!\!\!\!\!\!\!
    \varDelta_{n \to m} (x_m)
    &\!\!\!\!\!=\!\!\!\!\!&
    \int \! {p_{\mathsf{y}_n \mid \bm{\mathsf{x}}} 
    \!
    \left( y_n \mid \bm{x} \right) 
    \!\!
    \prod_{m' \neq m}^{M} \!\! { 
    \varDelta_{n \leftarrow m' } (x_{m'})
    }
    }
    \,
    \mathrm{d}\bm{x}_{\backslash m},
    \label{equ: spa_n_to_m}
    \\
    \!\!\!\!\!\!\!\!\!\!\!
    \varDelta_{n \leftarrow m} (x_m)
    &\!\!\!\!\!=\!\!\!\!\!&
    p_{\mathsf{x}_m} (x_m) 
    \cdot
    \prod_{n' \neq n}^{N}
    { \varDelta_{n' \to m}(x_m) },
    \label{equ: spa_m_to_n}
\end{eqnarray}
\end{subequations}
where $ \varDelta_{n \leftarrow m} (x_m) $ corresponds to the posterior information, and its mean and variance are denoted by $ \check{x}_{n, m} $ and  $ \check{v}_{n, m}
\triangleq \mathbb{E}_{\mathsf{x}_{n, m} \mid \check{\mathsf{x}}_{n, m} = \check{x}_{n, m} }
\bigl[ \bigl| x_{n, m} - \check{x}_{n, m} \bigr|^2 \bigr]
$, respectively.
In \eqref{equ: spa_n_to_m}, $ \varDelta_{n \to m} (x_m) $ is calculated based on all the beliefs excluding $ \varDelta_{n \leftarrow m} (x_m) $, which is the self-feedback from itself as shown in Fig. \ref{fig: fg_spa_fn}, and likewise, in \eqref{equ: spa_m_to_n}, $ \varDelta_{n \leftarrow m} (x_m) $ is calculated based on all the beliefs excluding $ \varDelta_{n \to m} (x_m) $ as shown in Fig. \ref{fig: fg_spa_vn}.
This is the well-known primitive self-noise suppression mechanism in \ac{SPA}.

Next, we will derive \ac{GaBP}, starting with the approximation of $\varDelta_{n \to m} (x_m)$ in \eqref{equ: spa_n_to_m}.
To that end, we define $z_n$ and $\varepsilon_{n, m}$ as
\begin{eqnarray}
\label{equ: an_condition}
&&
\!\!\!\!\!\!\!\!\!
\!\!\!\!\!\!\!\!\!
\overbrace{
\sum_{m=1}^{M} a_{n, m} x_m
}^{ \triangleq z_n  }
= 
\overbrace{
\sum_{m' \neq m}^{M} a_{n, m'} \left( x_m - \check{x}_{n, m'} \right)
}^{ \triangleq \varepsilon_{n, m} }
\nonumber
\\
&&
\qquad \qquad  \qquad 
+
\,
\sum_{m' \neq m}^{M} a_{n, m'} \check{x}_{n, m'}
+
a_{n, m} x_m.
\end{eqnarray}
Under the assumption that $M,N\gg 1$ and the statistical behavior of each error term $ (x_m - \check{x}_{n, m}) $ is independent of each other, $ \varepsilon_{n, m} $ can be approximated as a complex Gaussian random variable in conformity with \ac{CLT}.
As a result, the extrinsic joint belief in \eqref{equ: spa_n_to_m} can be expressed as
%
\begin{eqnarray}
\label{equ: joint_post_approx}
\!\!\!\!\!\!\!\!\!
&&
\!\!\!\!\!\!\!\!\!
    \prod_{m' \neq m}^{M}
     \varDelta_{n \leftarrow m'}(x_{m'}) 
     \nonumber
     \\
     \!\!\!\!\!\!\!\!\!\!\!\!
    &\propto&
     \!\!\!\!
    \exp{  \left[ \!- \frac{|z_n \! - \! a_{n, m} x_m \! - \! \sum_{m' \neq m}^{M} a_{n, m'} \check{x}_{n, m'} |^2}{ \sum_{m' \neq m}^{M}
|a_{n, m'}|^2 \check{v}_{n, m'} }  \right] }.
\end{eqnarray}
Substituting \eqref{equ: joint_post_approx} and 
$ p_{\mathsf{y}_n \mid \bm{\mathsf{x}}} 
\!
\left( y_n \mid \bm{x} \right)
=
e^{-\frac{|y_n - z_n|^2}{\sigma_{\mathrm{w}}^2} } / (\pi \sigma_{\mathrm{w}}^2) $ into \eqref{equ: spa_n_to_m} yields\footnote{
Note that multiple integral \ac{w.r.t.} $ x_1, \ldots, x_{m-1}, x_{m+1}, \ldots , x_M $ in \eqref{equ: spa_n_to_m} reduces to the integral \ac{w.r.t.} a single parameter $z_n$ owing to \ac{CLT}.
}
\begin{equation}
\label{equ: spa_n_to_m_approx}
\varDelta_{n \to m} (x_m)
\propto
\exp{\left[ 
-
\frac{|\tilde{y}_{n, m} - a_{n, m} x_m |^2 }{\psi_{n, m}}
\right]},
\end{equation}
where we employ the following predictive distribution rule for Gaussian distribution:
\begin{equation}
e^{ - \frac{ |a - b|^2 }{ v_1 + v_2 } } \propto \int_{\mathbb{C}} 
e^{ - \frac{ |a - x|^2 }{ v_1 } } 
\cdot
e^{ - \frac{ |x - b|^2 }{ v_2 } }  \, \mathrm{d}x,
\end{equation}
and then define 
\begin{subequations}
\begin{eqnarray}
    \tilde{y}_{n, m} \!\!\!\!&\triangleq&\!\!\!\! y_n - \sum_{m' \neq m}^{M} a_{n, m'} \check{x}_{n, m'},
    \\
    \psi_{n, m} \!\!\!\!&\triangleq&\!\!\!\!  \sum_{m' \neq m}^{M} |a_{n, m'}|^2 \check{v}_{n, m'}  + \sigma_{\mathrm{w}}^2.
\end{eqnarray}
\end{subequations}

Similarly, substituting \eqref{equ: spa_n_to_m_approx} into \eqref{equ: spa_m_to_n} yields
\begin{eqnarray}
\label{equ: spa_m_to_n_approx}
\varDelta_{n \leftarrow m} (x_m)
\!\!\!\!&\propto&\!\!\!\!
p_{\mathsf{x}_m} (x_m) 
\cdot
\exp{\left[ 
-
\sum_{n' \neq n}^{N}
\frac{|\tilde{y}_{n', m} - a_{n', m} x_m |^2 }{\psi_{n', m}}
\right]}
\nonumber
\\
\!\!\!\!&\propto&\!\!\!\!
p_{\mathsf{x}_m} (x_m) 
\cdot
\exp{\left[ 
-
\frac{\left|x_m - \overline{x}_{n, m} \right|^2}{\overline{v}_{n, m}}
\right]},
\end{eqnarray}
where we define
\begin{subequations}
\label{eq:GaBP_LEouput}
\begin{eqnarray}
    \overline{x}_{n, m}
    \!\!\!\!&\triangleq&\!\!\!\!
    \overline{v}_{n, m}
    \cdot
    \sum_{n' \neq  n}^{N} \frac{ a_{n', m}^* \tilde{y}_{n', m} }{\psi_{n', m}}, \\
    \overline{v}_{n, m}
    \!\!\!\!&\triangleq&\!\!\!\!
    \left( \sum_{n' \neq  n}^{N} \frac{\left| a_{n', m} \right|^2 }{ \psi_{n', m} } \right)^{-1}.
\end{eqnarray}
\end{subequations}

Finally, from \eqref{equ: spa_m_to_n_approx} and \eqref{eq:GaBP_LEouput}, the soft replica and its \ac{MSE}, \textit{i.e.}, the mean and variance of the posterior information, $\check{x}_{n, m}$ and $\check{v}_{n, m}$, can be obtained by the Bayesian optimal denoiser in \eqref{equ: bayes_optimal_denoiser} and \eqref{equ: denoiser_mse} as
\begin{subequations}
\begin{eqnarray}
    \!\!\!\!\!\!
    \check{x}_{n, m}
    \!\!\!\!&=&\!\!\!\!
    \frac{
    \displaystyle
    \int x_{m} \cdot \varDelta_{n \leftarrow m} (x_m) \, \mathrm{d} x_m }{
    \displaystyle
    \int \varDelta_{n \leftarrow m} (x_m') \, \mathrm{d} x_m'
    }
    =
    \eta \left( \overline{x}_{n, m} ; \overline{v}_{n, m} \right),
    \\
    \!\!\!\!\!\!
    \check{v}_{n, m} 
    \!\!\!\!&=&\!\!\!\!
    \overline{v}_{n, m} \cdot \frac{\partial \eta \left( \overline{x}_{n, m} ; \overline{v}_{n, m} \right)}{\partial \overline{x}_{n, m}}.
\end{eqnarray}
\end{subequations}
%
%
\begin{algorithm}[!t]
\hrulefill
\begin{algorithmic}[1]
\vspace{-0.5ex}
\Require{$\bm{y}\in \mathbb{C}^{N\times 1}, \bm{A} \in \mathbb{C}^{N \times M}, \, T$, \textbf{Output:} $\hat{x}_{m},  \forall m$}
    \\
    \STATE
    $ \forall (n, m) : \check{x}_{n, m}^{(1)} = 0, \, \check{v}_{n, m}^{(1)} = \sigma_{\mathrm{x}}^2 $
    \Comment{Initialization}
    \For{$t=1\ \mathrm{to}\ T$}
    \LineComment{/* LE */}
    \\
    \STATE 
    $
    \forall (n, m)
    : 
    \tilde{y}_{n, m}^{(t)} = y_n - \sum_{j \neq m}^{M} a_{n, j} \check{x}_{n, j}^{(t)} $
    \\
    \STATE 
    $
    \forall (n, m)
    :
    \psi_{n, m}^{(t)} = \sum_{j \neq m}^{M} 
    \left| a_{n, j} \right|^2
    \check{v}_{n, j}^{(t)} 
    + \sigma_{\mathrm{w}}^2
    $
    \Comment{IC}
    \\
    \STATE 
    $
    \forall (n, m)
    :    
    \overline{v}_{n, m}^{(t)}
    =
    \left( \sum_{i \neq n}^{N} \frac{\left| a_{i, m} \right|^2 }{ \psi_{i, m}^{(t)} } \right)^{-1}
    $
    \\
    \STATE 
    $
    \forall (n, m)
    :    
    \overline{x}_{n, m}^{(t)}
    =
    \overline{v}_{n, m}^{(t)} 
    \cdot
    \sum_{i \neq n}^{N} \frac{ a_{i, m}^* \tilde{y}_{i, m}^{(t)} }{\psi_{i, m}^{(t)}}
    $ 
    \Comment{MF}
    \LineComment{/* NLE  */} 
    \\
    \STATE 
    $
    \forall (n, m)
    :    
    \check{x}_{n, m}^{(t+1)}
    =
    \eta \left( \overline{x}_{n, m}^{(t)} ; \overline{v}_{n, m}^{(t)} \right)$
    \\
    \STATE 
    $
    \forall (n, m)
    :
    \check{v}_{n, m}^{(t+1)} =
    \overline{v}_{n, m}^{(t)} \cdot \frac{\partial \eta \left( \overline{x}_{n, m}^{(t)} ; \overline{v}_{n, m}^{(t)} \right)}{\partial \overline{x}_{n, m}}$
    \Comment{Denoising}
\EndFor
\\
\STATE 
$
\forall m
:    
\overline{v}_{m}
\!=\!
\left( \sum_{n = 1}^{N} \frac{\left| a_{n, m} \right|^2 }{ \psi_{n, m}^{(T)} } \right)^{\!\!-1}
\!\!\!\!\!\!,\  
\overline{x}_{m}
\!=\!
\overline{v}_{m}
\cdot
\sum_{n=1}^{N} \frac{ a_{n, m}^* \tilde{y}_{n, m}^{(T)} }{\psi_{n, m}^{(T)}}
$ 
\\
\STATE
$ \forall m :
 \hat{x}_m \!= 
\eta \left( \overline{x}_{m} ; \overline{v}_{m} \right)
$
\caption[]{- GaBP~\cite{Takahashi2019_tcom}, \textit{a.k.a.}, ABP~\cite{Kabashima2003}  }
\label{alg: gabp}
\end{algorithmic}
\end{algorithm}
\setlength{\textfloatsep}{5pt}
%

The pseudo code of \ac{GaBP} derived above is summarized in Algorithm~\ref{alg: gabp}, where $ (\cdot)^{(t)} $ denotes the corresponding variable at the $t$-th iteration step for all $ t \in \left\{1, 2, \cdots , T \right\} \triangleq \mathcal{T} $.
For ease of notation, the $i$-th line of Algorithm~\ref{alg: gabp} is referred
 to as (A\ref{alg: gabp}-$i$) hereafter.
In \ac{LE}, \ac{IC} is performed using the soft replica obtained in the previous iteration in (A\ref{alg: gabp}-3) and (A\ref{alg: gabp}-4), and then in (A\ref{alg: gabp}-5) and (A\ref{alg: gabp}-6), signal separation based on \ac{MF} and \textit{extrinsic belief combining} inherited from \ac{SPA} are performed.
In \ac{NLE}, by treating the \ac{LE} output as a virtual \ac{AWGN}-corrupted observation, the symbol-wise \ac{MMSE} estimate is computed using the denoiser of (A\ref{alg: gabp}-7) and (A\ref{alg: gabp}-8).

\vspace{-2mm}
\subsection{MF-EP}

\begin{figure*}[t]
\begin{center}
\subfloat[Belief from FN to VN]{
	\includegraphics[width=0.75\columnwidth,keepaspectratio=true]{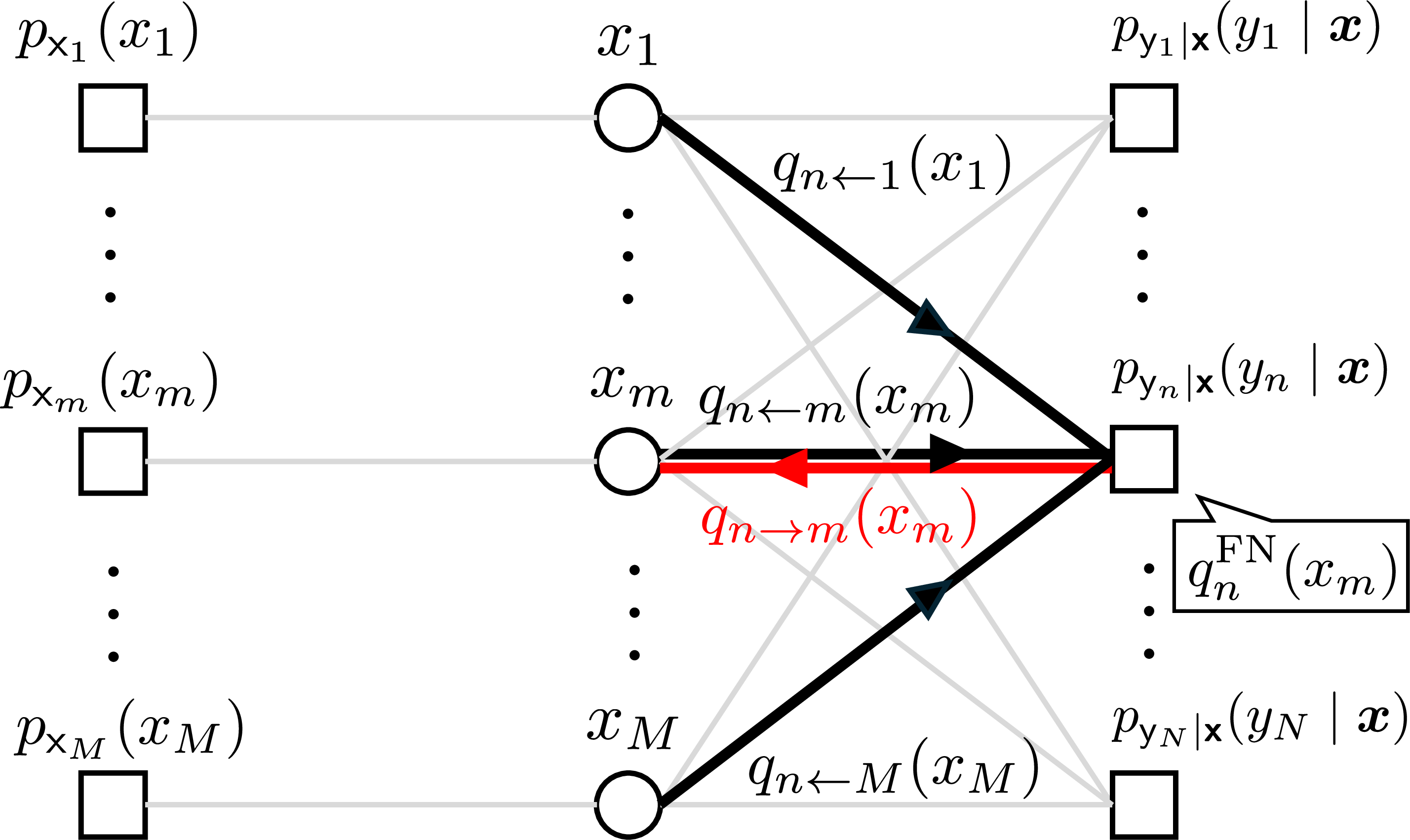}
	\label{fig: fg_mfep_fn}
	}
 	\hspace{0mm}
	\subfloat[Belief from VN to FN]{
	\includegraphics[width=0.75\columnwidth,keepaspectratio=true]{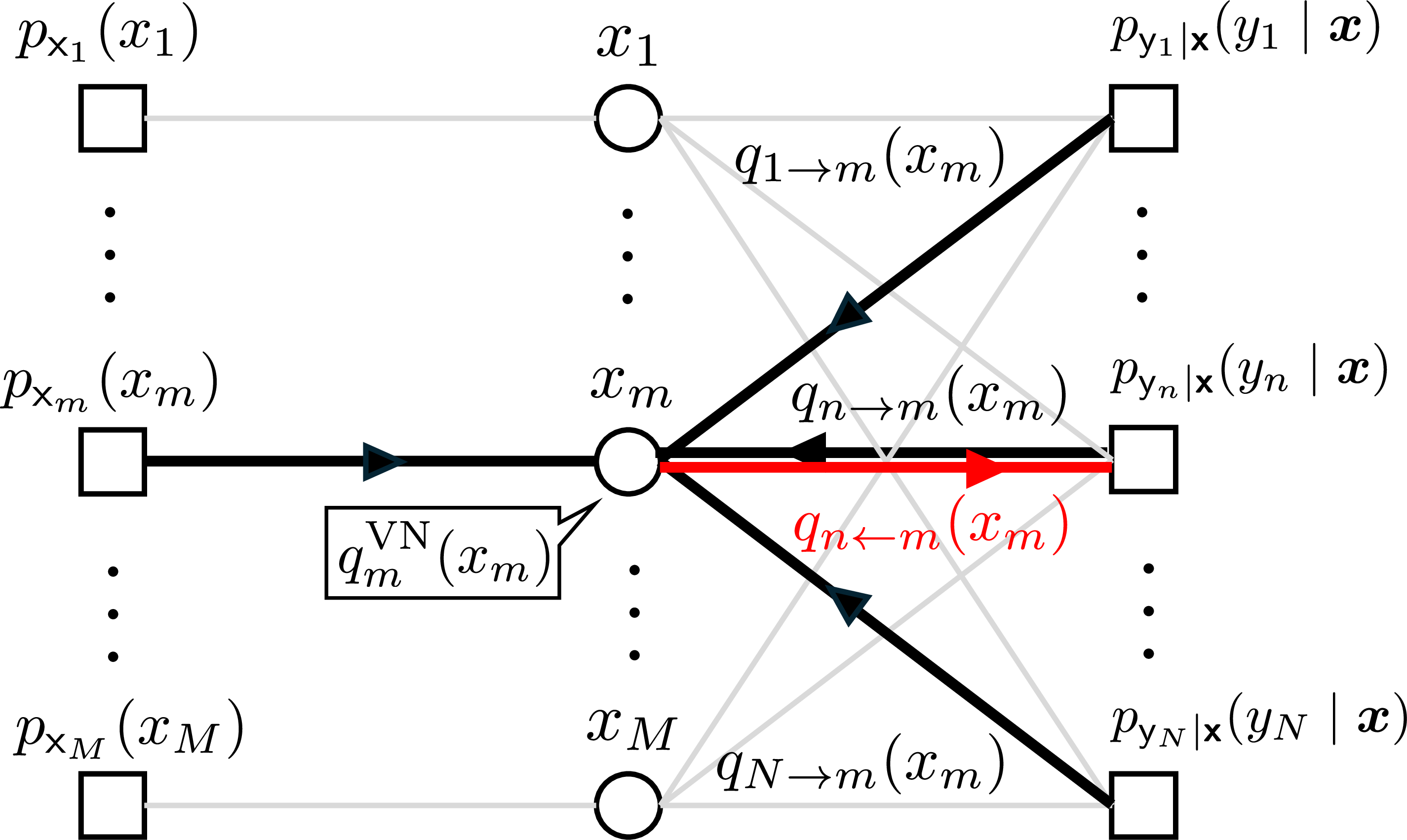}
	\label{fig: fg_mfep_vn}
	}
	\caption{The illustration of update rules of \ac{MF}-\ac{EP} on the \ac{FG}.}
	\label{fig: fg_mfep}
	\vspace{-5mm}
\end{center}
\end{figure*}

In this subsection, we briefly describe the algorithm of \ac{MF-EP}~\cite{Meng2015}, and clarify the self-noise suppression mechanism by comparing it with that of \ac{GaBP}. 
In \ac{MF}-\ac{EP}, the belief updates for the \ac{FN} and \ac{VN} shown in Figs. \ref{fig: fg_mfep_fn} and \ref{fig: fg_mfep_vn}, respectively, can be expressed as
\begin{subequations}
\begin{eqnarray}
\label{equ: q_n}
&&
\!\!\!\!\!\!\!\!\!\!\!\!\!\!\!\!\!\!
\!\!\!\!\!\!
q_{n}^{\mathrm{FN}} (x_m)
\nonumber
\\
&&
\!\!\!\!\!\!\!\!\!\!\!\!\!\!\!\!\!\!
\!\!\!\!\!\!
= 
\mathrm{Proj}_{\mathcal{F}}
\!
\left[
\int{p_{\mathsf{y}_n \mid \bm{\mathsf{x}}} 
\left( y_n \mid \bm{x} \right) 
\prod_{m' = 1}^{M}{
q_{n \leftarrow m' } (x_{m'})
}
}
\,
\mathrm{d}\bm{x}_{\backslash m}
\!
\right] \!\! ,
\\
&&
\label{equ: q_n_m}
\!\!\!\!\!\!\!\!\!\!\!\!\!\!\!\!\!\!
\!\!\!\!\!\!
q_{n \to m} (x_m)
    \propto
    \frac{q_{n}^{\mathrm{FN}} (x_m) }{ q_{n \leftarrow m} (x_m) },
\end{eqnarray}
\label{equ: mfep_fn}
\end{subequations}
and
\vspace{-2mm}
\begin{subequations}
\begin{eqnarray}
    \label{equ: q_m}
    \!\!\!\!\!\!\!\!\!\!\!\!\!\!\!\!\!\!
    q_{m}^{\mathrm{VN}} (x_m)
    &\!\!\!\!=\!\!\!\!&
    \mathrm{Proj}_{\mathcal{F}}
    \left[
    p_{\mathsf{x}_m} (x_m) 
    \cdot
    \prod_{n' = 1}^{N}
    { q_{{n'} \to m} (x_m) } 
    \right] \!\! ,
    \\
    \!\!\!\!\!\!\!\!\!\!\!\!\!\!\!\!\!\!
    \label{equ: q_m_n}
    q_{n \leftarrow m} (x_m)
    &\!\!\!\!\propto\!\!\!\!&
    \frac{q_{m}^{\mathrm{VN}} (x_m) }{ q_{n \to m} (x_m) },
\end{eqnarray}
\label{equ: mfep_vn}%
\end{subequations}
where $ \mathrm{Proj}_{\mathcal{F}} [\cdot] $ is an operator that projects the input distribution $p$ onto a tractable distribution $q \in \mathcal{F}$ such that the \ac{KLD} 
$ \mathcal{D}_{\mathrm{KL}} \left( p \parallel q \right) $ between $p$ and $q$ is minimized, \textit{i.e.}, 
\vspace{-0.5mm}
\begin{equation}
    \label{equ: kld_between_p_q}
    q
    =
    \underset{ q' \in \mathcal{F}  } {\operatorname{arg \, min}}
    \,
    \mathcal{D}_{\mathrm{KL}} \left( p \parallel q' \right).
\end{equation}
\vspace{-0.5mm}
In particular, when a set of distributions $\mathcal{F}$ is the exponential family, the minimization of \ac{KLD} in \eqref{equ: kld_between_p_q} can be achieved by \acf{MM}~\cite{Bishop2006}.

The crucial difference between \ac{GaBP} and \ac{MF-EP} lies in the method of generating extrinsic information.
In contrast to \ac{GaBP}, where each node merely excludes the belief from itself during belief combining, the \ac{MF-EP} first combines all beliefs, \textit{i.e.}, $q_{n \leftarrow m'}(x_m')$ and $q_{n' \to m}(x_m)$, at each node, as can be observed from \eqref{equ: q_n} and \eqref{equ: q_m}.
Subsequently, the combined belief is projected onto a tractable distribution, \textit{i.e.}, $ q_n(x_m) $ and $ q_m (x_m)$.
Finally, the self-feedback component is removed from the projected distribution through a division operation in the post-projection domain, as in \eqref{equ: q_n_m} and \eqref{equ: q_m_n}, such that the input and output beliefs of each node are uncorrelated.

The pseudo code of \ac{MF-EP} derived from \eqref{equ: mfep_fn} and \eqref{equ: mfep_vn} is given in Algorithm~\ref{alg: mfep}, where all the propagating messages are approximated as Gaussian distribution via \ac{MM}.
For further details, we refer the reader to~\cite{Meng2015}.
Interestingly, despite the fact that the underlying theory appears to be completely different, the only structural difference between Algorithms~\ref{alg: gabp} and~\ref{alg: mfep} is the self-noise suppression mechanism: extrinsic belief combining in (A\ref{alg: gabp}-5) and (A\ref{alg: gabp}-6), and \ac{MM} in (A\ref{alg: mfep}-9) and (A\ref{alg: mfep}-10) derived from \eqref{equ: q_m_n}.
In addition, the two algorithms have different relative positions of the noise suppression mechanism and the denoiser.
This observation in fact has important implications for the discussion in Section \ref{chap: performance}.


%
%
\begin{algorithm}[!t]
\hrulefill
\begin{algorithmic}[1]
\vspace{-0.5ex}
\Require{$\bm{y}\in \mathbb{C}^{N\times 1}, \bm{A} \in \mathbb{C}^{N \times M}, \, T$, \textbf{Output:} $\hat{x}_{m}^{(T)},  \forall m$}
    \\
    \STATE
    $ \forall (n, m) : \check{x}_{n, m}^{(1)} = 0, \, \check{v}_{n, m}^{(1)} = \sigma_{\mathrm{x}}^2 $
    \Comment{Initialization}
    \For{$t=1\ \mathrm{to}\ T$}
    \LineComment{/* LE */}
    \\
    \STATE 
    $
    \forall (n, m)
    : 
    \tilde{y}_{n, m}^{(t)} = y_n - \sum_{j \neq m}^{M} a_{n, j} \check{x}_{n, j}^{(t)} $
    \\
    \STATE 
    $
    \forall (n, m)
    :
    \psi_{n, m}^{(t)} = \sum_{j \neq m}^{M} 
    \left| a_{n, j} \right|^2 
    \check{v}_{n, j}^{(t)}
    + \sigma_{\mathrm{w}}^2
    $
    \Comment{IC}
    \\
    \STATE 
    $
    \forall m
    :    
    \overline{v}_{m}^{(t)}
    =
    \left( \sum_{n=1}^{N} \frac{\left| a_{n, m} \right|^2 }{ \psi_{n, m}^{(t)} } \right)^{-1} 
    $
    \\
    \STATE 
    $
    \forall m
    :    
    \overline{x}_{m}^{(t)}
    =
    \overline{v}_{m}^{(t)}
    \cdot
    \sum_{n=1}^{N} \frac{ a_{n, m}^* \tilde{y}_{n, m}^{(t)} }{\psi_{n, m}^{(t)}}
    $ 
    \Comment{MF}
    \LineComment{/* NLE  */} 
    \\
    \STATE 
    $
    \forall m
    :    
    \hat{x}_m^{(t)}
    =
    \eta \left( \overline{x}_{m}^{(t)} ; \overline{v}_{m}^{(t)} \right)$
    \\
    \STATE 
    $
    \forall m
    :
    \hat{v}_{m}^{(t)} =
    \overline{v}_{m}^{(t)} \cdot \frac{\partial \eta \left( \overline{x}_{m}^{(t)} ; \overline{v}_{m}^{(t)} \right)}{\partial \overline{x}_{m}}$
    \Comment{Denoising}
    %
    \\
    \STATE 
    $
    \forall (n, m)
    :    
    \frac{1}{\check{v}_{n, m}^{(t+1)}} = \frac{1}{\hat{v}_m^{(t)}} - \frac{|a_{n, m}|^2}{\psi_{n, m}^{(t)}} $
    \\
    \STATE 
    $
    \forall (n, m): \check{x}_{n, m}^{(t+1)}
    =
    \check{v}_{n, m}^{(t+1)}
    \cdot
    \left(
    \frac{\hat{x}_m^{(t)}}{\hat{v}_m^{(t)}}
    -
    \frac{a_{n, m}^* \tilde{y}_{n, m}^{(t)} }{ \psi_{n, m}^{(t)} }
    \right)$
    \Comment{MM}
    \EndFor
\caption[]{-  MF-EP~\cite{Meng2015}}
\label{alg: mfep}
\end{algorithmic}
\end{algorithm}
\setlength{\textfloatsep}{5pt}

\section{Derivation of GAMP from GaBP}
\label{chap: derivation_of_gamp}
In this section, we derive \ac{GAMP} from \ac{GaBP} by applying the large-system approximation to Algorithm~\ref{alg: gabp} and demonstrate that the self-noise suppression mechanism in \ac{GaBP} is equivalent to the Onsager correction term in \ac{GAMP} in the large-system limit.
Note that the following derivation process applies to an arbitrary prior distribution $ p_{\mathsf{x}_m}(\cdot) $.
Without loss of generality, each element in $\bm{A}$ independently follows an identical circularly symmetric distribution with mean zero and variance of order $1/N$.
In this section, let $\mathcal{O}(\cdot)$ denote the error term that vanishes in the large-system limit, \textit{i.e.}, as $N, M \to \infty$ with $\xi \triangleq N/M$ fixed.

\subsection{Approximation of Second-Order Moments}

We start the derivation by approximating the second-order moments in Algorithm~\ref{alg: gabp}.
First, the dependency of $ \psi_{n, m} $ in (A\ref{alg: gabp}-4) on $m$ is eliminated 
by neglecting the error term of order $ 1/N $ as
%
%
\vspace{-2mm}
\begin{eqnarray}
\label{equ: psi_approx1}
    \psi_{n, m}^{(t)}
    \!\!\!\!
    &=&
    \!\!\!\!
    \underbrace
    {
    \sum_{j=1}^{M} 
    \left| a_{n, j} \right|^2
    \check{v}_{n, j}^{(t)} 
    + \sigma_{\mathrm{w}}^2
    }_{\triangleq \psi_{n}^{(t)} }
    + \, \mathcal{O} \left(N^{-1}\right).
\end{eqnarray}
Similarly, the \ac{MSE} of the \ac{LE} output in (A\ref{alg: gabp}-5) can be expressed as
\vspace{-2mm}
\begin{eqnarray}
    \!\!\!\!\!\!\!\!
    \label{equ: overline_v_approx1}
    \overline{v}_{n, m}^{(t)}
    \!\!\!\!\!
    &=&
    \!\!\!\!\!
    \Biggl(
    \sum_{i \neq n}^{N} \frac{\left| a_{i, m} \right|^2 }{ 
    \psi_{i, m}^{(t)}
    }
    \Biggr)^{-1}
    \nonumber
    \\
    \!\!\!\!\!
    &\overset{\textrm{(a)}}{=}&
    \!\!\!\!\!
    \Biggl(
    \underbrace{
    \sum_{i=1}^{N} \frac{\left| a_{i, m} \right|^2 }{ \psi_{i}^{(t)} }
    }_{\triangleq \left( \overline{v}_{m}^{(t)} \right)^{-1} }
    +
    \,
    \mathcal{O} \left( N^{-1} \right)
    \Biggr)^{-1}  
    \!\!\!\!\!\!
    =
    \overline{v}_{m}^{(t)} + \mathcal{O} \! \left( N^{-1} \right) \! ,
\end{eqnarray}
where we use \eqref{equ: psi_approx1} in (a).

\vspace{-2mm}
\subsection{Approximation of First-Order Moments}
Likewise, we approximate the first-order moments in Algorithm~\ref{alg: gabp}.
For later convenience, we first rewrite the output of the \ac{IC} process as
\vspace{-2mm}
\begin{eqnarray}
    \label{equ: tilde_y_n_m_approx1}
    \tilde{y}_{n, m}^{(t)} 
    \!\!\!\!
    &=& 
    \!\!\!\!
    y_n
    -
    \underbrace
    {
    \sum_{j=1}^{M} a_{n, j} \check{x}_{n, j}^{(t)}
    }_{\triangleq p_{n}^{(t)} }
    \,
    +
    \,
    a_{n, m}
     \check{x}_{n, m}^{(t)}.
\end{eqnarray}
Using \eqref{equ: tilde_y_n_m_approx1}, the dependency of the estimate 
$ \overline{x}_{n, m}^{(t)} $
in (A\ref{alg: gabp}-6) on $n$ is eliminated as
\begin{eqnarray}
    \label{equ: overline_x_approx1}
    \overline{x}_{n, m}^{(t)}
    \!\!\!\!
    &\overset{\textrm{(b)}}{=}& 
    \!\!\!\!
    \overline{v}_{m}^{(t)}
    \sum_{i \neq n}^{N} \frac{ a_{i, m}^* \tilde{y}_{i, m}^{(t)} }{\psi_{i}^{(t)} }
    +
    \mathcal{O}(N^{-1})
    \nonumber
    \\
    \!\!\!\!
    &=& 
    \!\!\!\!  
    \overline{v}_{m}^{(t)} 
    \sum_{i=1}^{N} \frac{ a_{i, m}^* \tilde{y}_{i, m}^{(t)} }{\psi_{i}^{(t)}}
    -
    \overline{v}_{m}^{(t)}    
    \cdot    
    \frac{a_{n, m}^* \tilde{y}_{n, m}^{(t)}}{\psi_n^{(t)}}
    +
    \mathcal{O}\left(N^{-1} \right)  
    \nonumber
    \\
    \!\!\!\!
    &\overset{\textrm{(c)}}{=}& 
    \!\!\!\! 
    \overline{v}_{m}^{(t)} 
    \sum_{i=1}^{N} 
    a_{i, m}^* \cdot
     \overbrace
     {
    \frac{
     y_{i} - p_{i}^{(t)} }{\psi_{i}^{(t)}}
    }
    ^{\triangleq s_{i}^{(t)} }
    +
    \,
    \overline{v}_{m}^{(t)}
    \sum_{i=1}^{N}
    \frac{
    \left| a_{i, m} \right|^2
    }{ \psi_{i}^{(t)} }
    \cdot
    \check{x}_{i, m}^{(t)}
    \nonumber
    \\
    &&   
    \qquad\qquad\qquad\qquad
    -
    \,\,
    \overline{v}_{m}^{(t)}    
    \cdot
    \frac{a_{n, m}^* \tilde{y}_{n, m}^{(t)}}{\psi_n^{(t)}}
    +
    \mathcal{O}\left( N^{-1} \right)    
    \nonumber
    \\
    \!\!\!\!
    &\overset{\textrm{(d)}}{=}& 
    \!\!\!\!
    \label{equ: belief_mean_approx_1}    
    \overline{x}_{m}^{(t)}
    -
    \underbrace{
    \overline{v}_{m}^{(t)}    
    \cdot    
    \frac{a_{n, m}^* \tilde{y}_{n, m}^{(t)}}{\psi_n^{(t)}}
    }_{\triangleq \Delta \overline{x}_{n, m}^{(t)} }
    +
    \,
    \mathcal{O}\left( N^{-1} \right),  
\end{eqnarray}
where \eqref{equ: psi_approx1} and \eqref{equ: overline_v_approx1} are used in (b), and \eqref{equ: tilde_y_n_m_approx1} is applied  in (c).
In addition, we define $ \overline{x}_m^{(t)} $ in (d) as 
\begin{equation}
\label{equ: overline_x_m}
\overline{x}_m^{(t)}
\triangleq
\overline{v}_m^{(t)}
\sum_{n=1}^{N}
\frac{ a_{n, m}^* s_n^{(t)} }{\psi_n^{(t)}}
+
\overline{v}_m^{(t)}
\sum_{n=1}^{N}
\frac{|a_{n, m}|^2 \check{x}_{n, m}^{(t)}}{\psi_n^{(t)}}.
\end{equation}
Using \eqref{equ: tilde_y_n_m_approx1}, the term $ \Delta \overline{x}_{n, m}^{(t)} $ in \eqref{equ: overline_x_approx1}, \textit{i.e.},
 the self-feedback component, can be further rewritten as
\begin{eqnarray}
    \label{equ: def_Delta_x}
    \Delta \overline{x}_{n, m}^{(t)}    
    \!\!\!\!
    &=&
    \!\!\!\!  
    \overline{v}_{m}^{(t)}    
    \cdot    
    a_{n, m}^*
    s_n^{(t)}
    +
    \mathcal{O} \left( N^{-1} \right).
\end{eqnarray}
\subsection{Approximation of Soft Replica and its MSE}
Next, we evaluate the effect of the self-noise feedback
$ \Delta \overline{x}_{n, m}^{(t)} $
propagating to the next iteration via the Bayesian optimal denoiser in (A\ref{alg: gabp}-7) and (A\ref{alg: gabp}-8) using the Taylor expansion.
First, from the chain rule of Wirtinger derivative, we have
\begin{equation}
\label{equ: chain_rule}
    \frac{\partial}{\partial \overline{x}_{n, m}}
    =
    \frac{\partial }{\partial \overline{x}_{m}}
    \cdot
    \!\!
    \underbrace{
    \frac{\partial \overline{x}_{m}}{\partial \overline{x}_{n, m}}
    }_{=1 \textrm{ from \eqref{equ: overline_x_approx1} }}
    \!\!
    +
    \,
    \frac{\partial }{\partial \overline{x}_{m}^*}
    \cdot
    \!\!
    \underbrace{
    \frac{\partial \overline{x}_{m}^*}{\partial \overline{x}_{n, m}}
    }_{=0 \textrm{ from \eqref{equ: overline_x_approx1} }}
    \!\!
    =
    \frac{\partial }{\partial \overline{x}_{m}}.
\end{equation}
Substituting \eqref{equ: overline_v_approx1}, \eqref{equ: overline_x_approx1}, and \eqref{equ: chain_rule} into (A\ref{alg: gabp}-8) yields
\begin{eqnarray}
    \label{equ: check_v_approx}
    \check{v}_{n, m}^{(t+1)}
    \!\!\!\!
    &=& 
    \!\!\!\!
    \overline{v}_{m}^{(t)} 
    \cdot \frac{\partial }{\partial \overline{x}_{m}}   
    \eta 
    \left(
    \overline{x}_m^{(t)}
    -
    \Delta \overline{x}_{n, m}^{(t)}
    ; \overline{v}_{m}^{(t)}
    \right)
    +
    \mathcal{O} \left( N^{-1} \right)
    \nonumber
    \\
    \!\!\!\!
    &=& 
    \!\!\!\!
    \underbrace{
    \overline{v}_{m}^{(t)} 
    \cdot
    \frac{\partial }{\partial \overline{x}_{m}}
    \eta 
    \left(
    \overline{x}_m^{(t)}
    ; \overline{v}_{m}^{(t)}
    \right)
    }_{\triangleq \check{v}_{m}^{(t+1)} }
    + \,
    \mathcal{O}
    \left(
    N^{-\frac{1}{2}}
    \right),
\end{eqnarray}
which follows from $ \Delta \overline{x}_{n, m}^{(t)} = \mathcal{O} \left(N^{-1/2} \right)$.

Similarly, from (A\ref{alg: gabp}-7) we have 
\begin{eqnarray}
    \label{equ: check_x_approx}
    \check{x}_{n, m}^{(t+1)}
    \!\!\!\!
    &=& 
    \!\!\!\!
    \eta \! \left( \overline{x}_{n, m}^{(t)} ; \overline{v}_{n, m}^{(t)} \right)
    =
    \eta \! \left( \overline{x}_{m}^{(t)} -
    \Delta \overline{x}_{n, m}^{(t)}
     ; \overline{v}_{m}^{(t)}
     \right) 
     +
     \mathcal{O} (N^{-1} )
    \nonumber
    \\
    \!\!\!\!
    &\overset{\textrm{(e)}}{=}& 
    \!\!\!\!
    \overbrace{
    \eta \! \left( \overline{x}_{m}^{(t)}  
     ; \overline{v}_{m}^{(t)}
     \right) 
     }^{\triangleq \check{x}_{m}^{(t+1)} }
    -
    \Delta \overline{x}_{n, m}^{(t)}
    \! \cdot \!
    \frac{\partial }{\partial \overline{x}_{m}}
    \eta 
    \!
    \left(
    \overline{x}_m^{(t)}
    ; \overline{v}_{m}^{(t)}
    \right)
    \!
    \nonumber
    \\
    &&
    \qquad
    -
    \left(\Delta \overline{x}_{n, m}^{(t)}\right)^*
    \! \cdot \!
    \frac{\partial }{\partial \overline{x}_{m}^*}
    \eta 
    \!
    \left(
    \overline{x}_m^{(t)}
    ; \overline{v}_{m}^{(t)}
    \right)
    \!
    \nonumber
    +
    \mathcal{O} (N^{-1} )
    \nonumber
    \\
    \!\!\!\!
    &\overset{\textrm{(f)}}{=}& 
    \!\!\!\!
    \check{x}_m^{(t+1)}
    -
    a_{n, m}^*
    s_n^{(t)}
    \overbrace{
    \overline{v}_{m}^{(t)}
    \cdot
    \frac{\partial }{\partial \overline{x}_{m}}
    \eta 
    \left(
    \overline{x}_m^{(t)}
    ; \overline{v}_{m}^{(t)}
    \right)
    }^{= \check{v}_{m}^{(t+1)} } 
    \nonumber
    \\
    &&
    \quad
    -
    \,
   a_{n, m} \bigl(s_n^{(t)} \bigr)^*
   \underbrace{
   \overline{v}_{m}^{(t)}
    \! \cdot \!
    \frac{\partial }{\partial \overline{x}_{m}^*}
    \eta 
    \!
    \left(
    \overline{x}_m^{(t)}
    ; \overline{v}_{m}^{(t)}
    \right)
    }_{\triangleq \check{r}_{m}^{(t+1)} }
    \!
    \nonumber
    +
    \,
    \mathcal{O} (N^{-1} )
    \nonumber
    \nonumber
    \\
    \!\!\!\!
    &=& 
    \!\!\!\!    
    \check{x}_m^{(t+1)}
    -
    a_{n, m}^*
    s_n^{(t)}
    \check{v}_{m}^{(t+1)}
    \nonumber
    \\
    &&
    \qquad \qquad
    -
    \,
    a_{n, m}
    \bigl( s_n^{(t)} \bigr)^*
    \check{r}_{m}^{(t+1)}
    +
    \mathcal{O} (N^{-1} ),
\end{eqnarray}
where we use the Taylor expansion in (e) and \eqref{equ: def_Delta_x} in (f).
Here, the terms 
$ a_{n, m}^* s_n^{(t)} \check{v}_{m}^{(t+1)} +
a_{n, m}
\bigl( s_n^{(t)} \bigr)^*
\check{r}_{m}^{(t+1)}$
represent the self-feedback component contained in the output of the Bayesian optimal denoiser.
\subsection{Closing Loop}
Finally, we substitute \eqref{equ: check_v_approx} and \eqref{equ: check_x_approx} into
the definitions of $ \psi_n^{(t)}, \overline{x}_m^{(t)} $, and $ p_n^{(t)} $
to close the loop of iterative process.
From \eqref{equ: psi_approx1} and \eqref{equ: overline_x_m},
\begin{eqnarray}
    \psi_{n}^{(t)}
    \!\!\!\!
    &=&
    \!\!\!\!
    \underbrace
    {
    \sum_{m=1}^{M} 
    \left| a_{n, m} \right|^2
    \check{v}_{m}^{(t)}
    }_{\triangleq \gamma_{n}^{(t)} }
    +
    \,
    \sigma_{\mathrm{w}}^2
    +
    \mathcal{O} \left( N^{-\frac{1}{2}} \right),
\\
    \overline{x}_m^{(t)}
    \!\!\!\!
    &=&
    \!\!\!\!
    \overline{v}_{m}^{(t)}
    \sum_{n=1}^{N} 
     a_{n, m}^* 
    s_{n}^{(t)}
    \nonumber
    \\
    &&
    +
    \,
    \overline{v}_{m}^{(t)}
    \underbrace
    {
    \left(
    \sum_{n=1}^{N}
    \frac{
    \left| a_{n, m} \right|^2
    }{ \psi_{n}^{(t)} }
    \right)
    }_{= \left( \overline{v}_{m}^{(t)} \right)^{-1}} 
    \Biggl(
    \check{x}_{m}^{(t)}
    +
    \mathcal{O}\left( N^{-\frac{1}{2}} \right)
    \Biggr)
    \nonumber
    \\
    \!\!\!\!
    &=&
    \!\!\!\!
    \check{x}_{m}^{(t)}
    +
    \overline{v}_{m}^{(t)} 
    \sum_{n=1}^{N} 
     a_{n, m}^* 
    s_{n}^{(t)}
    +
    \mathcal{O} \left( N^{-\frac{1}{2}} \right).
\end{eqnarray}
In addition, from \eqref{equ: tilde_y_n_m_approx1}, we have
\begin{eqnarray}
    p_{n}^{(t)}
    \!\!\!\!\!
    &\overset{\textrm{(g)}}{=}&
    \!\!\!\!\!
    \sum_{m=1}^{M} a_{n, m} 
    \left(
    \check{x}_{m}^{(t)}
    -
    \check{v}_m^{(t)} 
    a_{n, m}^*
    s_n^{(t-1)}
    \right)
    +
    \mathcal{O}\left( N^{-\frac{1}{2}} \right)
    \nonumber
    \\
    \!\!\!\!\!
    &=&
    \!\!\!\!\!
    \sum_{m=1}^{M} a_{n, m} \check{x}_{m}^{(t)}
    \! - \! 
    \underbrace{
    \left( 
    \sum_{m=1}^{M} \left| a_{n, m} \right|^2 \check{v}_m^{(t)}
    \!
    \right)
    }_{=\gamma_n^{(t)} }
    \!
    s_n^{(t-1)}
    + 
    \mathcal{O}  \left(  N^{-\frac{1}{2}} \right) 
    \nonumber
    \\
    \!\!\!\!\!
    &=&
    \!\!\!\!\!
    \sum_{m=1}^{M} a_{n, m} \check{x}_{m}^{(t)}
    -
    \!\!\!\!
    \underbrace{
    \gamma_n^{(t)}
    s_n^{(t-1)}
    }_{\textrm{\textit{Onsager correction}}}
    \!\!\!\!
    +
    \,
    \mathcal{O} \left( N^{-\frac{1}{2}} \right),
\end{eqnarray}
where $ \gamma_n^{(t)} s_n^{(t-1)} $ is called the Onsager correction term.
In (g), the term 
 $ 
 \bigl( s_n^{(t-1)} \bigr)^*
 \sum_{m=1}^{M}  a_{n, m}^2 \check{r}_m^{(t)} 
 $ is of order $1/\sqrt{N}$ and thus neglected based on the following lemma.
\begin{Lemma}
\label{lem: complex_feedback_0}
Suppose each entry of $\bm{A} \in \mathbb{C}^{N \times M}$ is \ac{i.i.d.} and follows a circularly symmetric
probabilistic distribution with mean zero and variance of order $1/N$, the following equality holds:
\begin{equation}
 \mathbb{E}_{\bm{\mathsf{A}}}
 \left[ \left| 
 \sum_{m=1}^{M}  a_{n, m}^2 \check{r}_m
 \right|^2 
 \right]
 = \mathcal{O}\left(N^{-1}\right),
\end{equation}
under the heuristic assumption that $ \left|\check{r}_{m}\right| $ is bounded for all $ m \in \left\{1, 2, \cdots , M \right\} $.
\end{Lemma}
\begin{IEEEproof}
    See Appendix \ref{chap: proof_of_lemma}.
\end{IEEEproof}

The pseudo code of \ac{GAMP} based on the asymptotic convergence of the remaining higher-order terms to zero with $M,N\to \infty$ is given in Algorithm~\ref{alg: gamp}.
%
%
\begin{algorithm}[!t]
\hrulefill
\begin{algorithmic}[1]
\vspace{-0.5ex}
\Require{$\bm{y}\in \mathbb{C}^{N\times 1}, \bm{A} \in \mathbb{C}^{N \times M}, \, T$, \textbf{Output:} $\check{x}_{m}^{(T+1)},  \forall m$}
    \\
    \STATE
    $ \forall m : \check{x}_{m}^{(1)} = 0, \, \check{v}_{m}^{(1)} = \sigma_{\mathrm{x}}^2 $
    \\
    \STATE
    $ \forall n : s_{n}^{(0)} = 0$  
    \Comment{Initialization}
    \For{$t=1\ \mathrm{to}\ T$}
    \LineComment{/* LE */}
    \\
    \STATE 
    $
    \forall n
    : 
    \gamma_n^{(t)}
    =
    \sum_{m=1}^{M} 
    \left| a_{n, m} \right|^2
    \check{v}_{m}^{(t)} $
    \\
    \STATE 
    $
    \forall n
    :
    p_n^{(t)}
    =
    \sum_{m=1}^{M} a_{n, m} \check{x}_{m}^{(t)}
    -
    \gamma_n^{(t)}
    s_n^{(t-1)}
    $
    \Comment{Onsager Correction}
    \\
    \STATE 
    $
    \forall m
    :
    \psi_n^{(t)}
    =
    \gamma_n^{(t)} + \sigma_{\mathrm{w}}^2
    $
    \\
    \STATE 
    $
    \forall m
    :
    s_n^{(t)}
    =
    \frac{
    y_n - p_n^{(t)}
    }{ \psi_{n}^{(t)} }
    $
    \Comment{IC}
    \\
    \STATE 
    $
    \forall m
    :    
    \overline{v}_m^{(t)}
    =
    \left(
    \sum_{n=1}^{N}
    \frac{
    \left| a_{n, m} \right|^2
    }{ \psi_{n}^{(t)} }
    \right)^{-1}
    $
    \\
    \STATE 
    $
    \forall m
    :    
    \overline{x}_m^{(t)}
    =
    \check{x}_m^{(t)}
    +
    \overline{v}_{m}^{(t)}
    \sum_{n=1}^{N} 
     a_{n, m}^* 
    s_{n}^{(t)}
    $ 
    \Comment{MF}
    \LineComment{/* NLE  */} 
    \\
    \STATE 
    $
    \forall m
    :    
    \check{x}_{m}^{(t+1)}
    =
    \eta \left( \overline{x}_{m}^{(t)} ; \overline{v}_{m}^{(t)} \right)
    $
    \\
    \STATE 
    $
    \forall m
    :
    \displaystyle
    \check{v}_{m}^{(t+1)} =
    \overline{v}_{m}^{(t)} \cdot \frac{\partial \eta \left( \overline{x}_{m}^{(t)} ; \overline{v}_{m}^{(t)} \right)}{\partial \overline{x}_{m}}
    $
    \Comment{Denoising}
    \EndFor
\caption[]{-  GAMP~\cite{Rangan2011} }
\label{alg: gamp}
\end{algorithmic}
\end{algorithm}
\setlength{\textfloatsep}{5pt}
As described above, the Onsager correction term in (A\ref{alg: gamp}-5) that serves as the self-noise suppression mechanism of \ac{GAMP} is derived from the extrinsic belief combining in \ac{GaBP}, and it cancels out the self-feedback component contained in the output of the Bayesian optimal denoiser, thereby enabling to decouple each iterative process of \ac{GAMP}.

\section{Annealed Discrete Denoiser for Discrete-Valued Signal Estimation}

\begin{figure*}[t]
\begin{center}
\subfloat[ $ \mathcal{X}_4 = \left\{ c_1 + \mathrm{j} c_2 \relmiddle| c_1, c_2 \in \left\{ -c, +c \right\} \right\}, \, c = \sqrt{\Es/2}.  $ ]{
	\includegraphics[width=0.90\columnwidth,keepaspectratio=true]{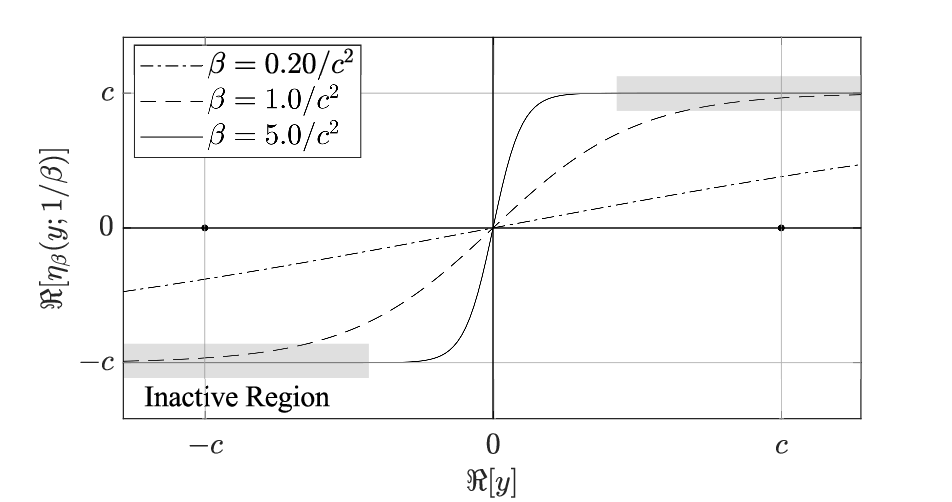}
	\label{fig: disc_denoiser_4qam}
	}
 	\hspace{0mm}
	\subfloat[ $ \mathcal{X}_{16} = \left\{ c_1 + \mathrm{j} c_2 \relmiddle| c_1, c_2 \in \left\{ \pm c, \pm 3c \right\} \right\}, \, c = \sqrt{\Es/10}. $]{
	\includegraphics[width=0.90\columnwidth,keepaspectratio=true]{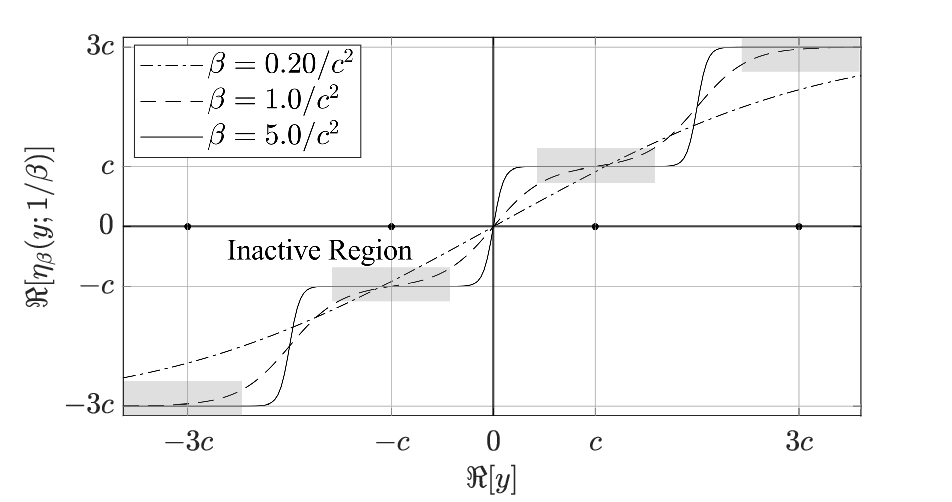}
	\label{fig: disc_denoiser_16qam}
	}
	\caption{The dynamics of the \ac{ADD} $ 
\eta_\beta (y; 1/\beta) $ for \ac{QAM} symbols with different values of $\beta$.}
\label{fig: disc_denoiser}
	\vspace{-6mm}
\end{center}
\end{figure*}

In this section, we generalize the belief scaling method, which was proposed in the context of \ac{GaBP}-based \ac{MIMO} wireless signal detection~\cite{Takahashi2019_tcom} as a method for improving its convergence property under non-ideal observation conditions. 
This generalization can be achieved by reinterpreting the operations performed on the belief of \ac{GaBP} in~\cite{Ito2024,Takahashi2023,Takahashi2022,Iimori2021} from the perspective of the Bayesian optimal denoiser.

Substituting the discrete prior in \eqref{equ: p_xm_w_disc_prior} into the Bayesian optimal denoiser in \eqref{equ: bayes_optimal_denoiser} yields
\begin{eqnarray}
    \label{equ: discrete_denoiser}
    \!\!\!\!\!\!\!\!\!\!\!\!
    &&
    \eta (y; v)=
    \sum_{\chi \in \mathcal{X}} \chi \cdot  \frac{P_{\mathsf{x}_m}[\chi] e^{- \frac{\left| y - \chi \right|^2}{ v }}
    }{
    \sum_{\chi' \in \mathcal{X}} P_{\mathsf{x}_m}[\chi'] e^{- \frac{\left| y - \chi' \right|^2}{ v }}}
    \nonumber \\
    \!\!\!\!\!\!\!\!\!\!\!\!
    &&\quad=
    \sum_{\chi \in \mathcal{X}} \chi \cdot  \frac{e^{\alpha(y;v)}}{
    \sum_{\chi' \in \mathcal{X}} e^{\alpha(y;v)}} =
    \sum_{\chi \in \mathcal{X}} \chi \cdot \zeta\left( \alpha(y; v) \right),
\end{eqnarray}
where $\zeta(\cdot)$ is the softmax function and for ease of notation we define the log posterior \ac{PDF} as
\begin{equation}
\label{eq:logpp}
    \alpha(y;v) \triangleq \ln P_{\mathsf{x}_m}[\chi] - \frac{\left| y - \chi \right|^2}{ v }.
\end{equation}
%

%


Under  conditions of (nearly ideal) large-scale uncorrelated observations, the optimal parameter $v$ is given analytically at each iteration step as the variance of the \ac{LE} output.
However, when the discrepancy between the idealized condition and actual system condition becomes significant, such as the finiteness of the system size and correlation among the elements of $\bm{A}$, the estimated variance, \textit{i.e.}, $ \overline{v}_{n, m}^{(t)} $ in  \textit(A\ref{alg: gabp}-5) and $ \overline{v}_{m}^{(t)} $ in (A\ref{alg: mfep}-5) and (A\ref{alg: gamp}-8), will no longer reflect the actual reliability of the belief due to the presence of model errors.
Indeed, especially in the early iterations, the denoiser in \eqref{equ: discrete_denoiser} that operates based on $ \overline{v}_{n, m}^{(t)} $ and $ \overline{v}_{ m}^{(t)} $ overestimates the reliability of the input belief.
Its inappropriate nonlinearity generates an incorrect hard-decision estimate that causes error propagation in the subsequent iterative \ac{IC} process, resulting in severe performance degradation.

Motivated by the above discussion, the \ac{ASB} is introduced in~\cite{Takahashi2019_tcom} according to the instantaneous channel state, but in fact, this operation is essentially equivalent to introducing the inverse temperature parameter $\beta^{(t)}$ in the softmax function instead of the unreliable $ \overline{v}_{n, m}^{(t)} $ or $ \overline{v}_{ m}^{(t)} $ as $v$ in \eqref{eq:logpp}.
In other words, the belief scaling process generalized for discrete-valued signal estimation is equivalent to replacing each denoising process, \textit{i.e.}, (A\ref{alg: gabp}-7, 8) of \ac{GaBP}, of (A\ref{alg: mfep}-7, 8) of \ac{MF-EP}, and (A\ref{alg: gamp}-10, 11) of \ac{GAMP}, by
the following parameterized denoiser:
\begin{eqnarray}
\label{equ: denoiser_w_asb}
    \eta_{\beta} \left( y ; 1/\beta \right)
    \!\!\!\!&\triangleq&\!\!\!\!
    \sum_{\chi \in \mathcal{X}} \chi \cdot \zeta\left( \alpha(y; 1/\beta) \right)
    ,
    \\
    \frac{1}{\beta} \cdot \frac{\partial \eta_{\beta}(y; 1/\beta)}{\partial y}
    \!\!\!\!&\triangleq&\!\!\!\!
    \sum_{\chi \in \mathcal{X}}
    \left|\chi\right|^2
    \zeta\left( \alpha(y; 1/\beta) \right)
    -
    \left|
    \eta_\beta \left( y ; 1/\beta \right)
    \right|^2. \nonumber
\end{eqnarray}
For later convenience, the denoiser in \eqref{equ: denoiser_w_asb} is hereafter referred to as an \textit{\acf{ADD}}.
The set of parameters $ \left\{ \beta^{(t)}; t \in \mathcal{T} \right\}$ should be designed empirically \cite{Takahashi2019_tcom} or via the use of learning optimization \cite{Shirase2020,Shirase2021} in such a way that the reliability of beliefs expected to be achieved at each iteration should be reflected in the dynamics of \eqref{equ: discrete_denoiser}.

An intuitive explanation of the relationship between the value of $ \beta$ and the shape of the denoiser in \eqref{equ: denoiser_w_asb} is given in Section VI using specific examples.

\section{Performance Assessment}
\label{chap: performance}
%
Numerical simulations were conducted to evaluate the performance of discrete-valued signal estimation
assuming the spatially-multiplexed \ac{MU-MIMO} signal detection problem in wireless communications.

\subsection{System Model}

Consider an uplink \ac{MU-MIMO} system consisting of $M$ \ac{UE} devices and one \ac{BS}, where each \ac{UE} device is equipped with a single \ac{TX} antenna while the \ac{BS} is equipped with $N$ \ac{RX} antennas in a \ac{ULA} pattern.
In this setting, $ x_m $ in \eqref{equ: linear_measurement} represents a \ac{TX} symbol from the $m$-th \ac{UE}
uniformly chosen from the set of Gray-coded $Q$-\ac{QAM} constellation points:
\begin{equation}
    \label{equ: chi_qam}
    \!\!\!\!\!
    \mathcal{X}_Q
    \!
    \triangleq
    \!
    \left\{ c_1 + \mathrm{j} c_2
    \!
    \relmiddle|
    \!
    c_1, c_2 \!\in \left\{ \pm c, \pm 3c, \! \cdots\!, \pm \bigl(\! \sqrt{Q}-1\!\bigr)c\right\}
    \right\}\!,
\end{equation}
with $c$ being set to
$\sqrt{3 \Es / \left( 2 (Q-1) \right)}$,
while $\bm{w}$ in \eqref{equ: linear_measurement} denotes the \ac{AWGN} vector of the noise power density $N_0 \left(= \sigma_{\mathrm{w}}^2 \right)$.
Based on the assumptions, $P_{\mathsf{x}_m}[\chi_q] = 1/Q, \, \forall \chi_q \in \mathcal{X}_Q, \, \forall m \in \left\{1, 2, \cdots , M \right\} $ holds.
In addition, the measurement matrix $\bm{A}$ in \eqref{equ: linear_measurement} denotes an \ac{MU-MIMO} channel matrix, whose $(n, m)$-th element $ a_{n, m} $ corresponds to the fading coefficient of the channel between the $m$-th \ac{UE} device and the $n$-th antenna element at the \ac{BS}, and we employ the typical Kronecker model~\cite{Chockalingam2014} to construct $\bm{A}$ as 
\begin{equation}
    \label{equ: channel_matrix}
    \bm{A} = \bm{R}_{\mathrm{RX}}^{1/2} \bm{G} \bm{R}_{\mathrm{TX}}^{1/2},
\end{equation}
where each element of $\bm{G} \in \mathbb{C}^{N \times M}$ representing the small-scale fading of the channel follows the \ac{i.i.d.} complex Gaussian distribution $\mathcal{CN} (0, 1)$, and $ \bm{R}_{\mathrm{TX}} \in \mathbb{C}^{M \times M}$ and $ \bm{R}_{\mathrm{RX}} \in \mathbb{C}^{N \times N} $ respectively
 denote the spatial correlation matrices on the \ac{TX} and \ac{RX} sides, though $\bm{R}_{\mathrm{TX}}$ is set to $\bm{I}_{M}$ in uplink \ac{MU-MIMO} systems.
For the purpose of evaluating the benchmark performance, each element of  $\bm{R}_{\mathrm{RX}}$ is generated based on the exponential attenuation model~\cite{Chiani2003} by
\begin{equation}
    \label{equ: R_rx_exponentially_decaying}
    \left[ \bm{R}_{\mathrm{RX}} \right]_{i, j}
    =
    \begin{cases}
    1, & i = j, \\
    \rho^{|i-j|}, & i \neq j,
    \end{cases}
\end{equation}
where $\rho \in [0, 1]$ denotes a fading correlation coefficient between two distinct \ac{RX} antennas on the \ac{BS} side.


\begin{figure*}[t]
\begin{center}
\subfloat[$ (M, N) = (16, 32), \, Q = 4, \, \rho = 0.90 $.]{
	\includegraphics[width=0.98\columnwidth,keepaspectratio=true]{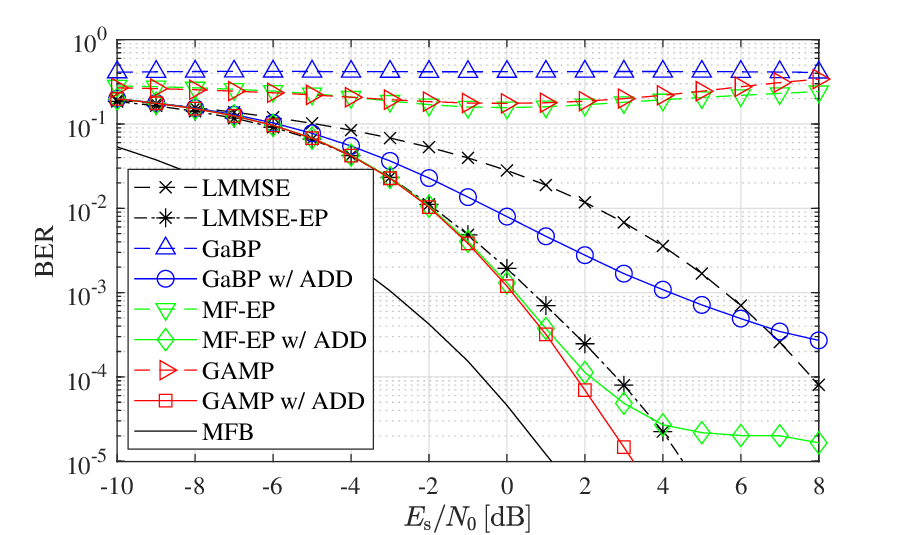}
	\label{fig: ber_16_32_4qam}
	}
 	\hspace{0mm}
	\subfloat[$ (M, N) = (16, 32), \, Q = 16, \, \rho = 0.80 $.]{
	\includegraphics[width=0.98\columnwidth,keepaspectratio=true]{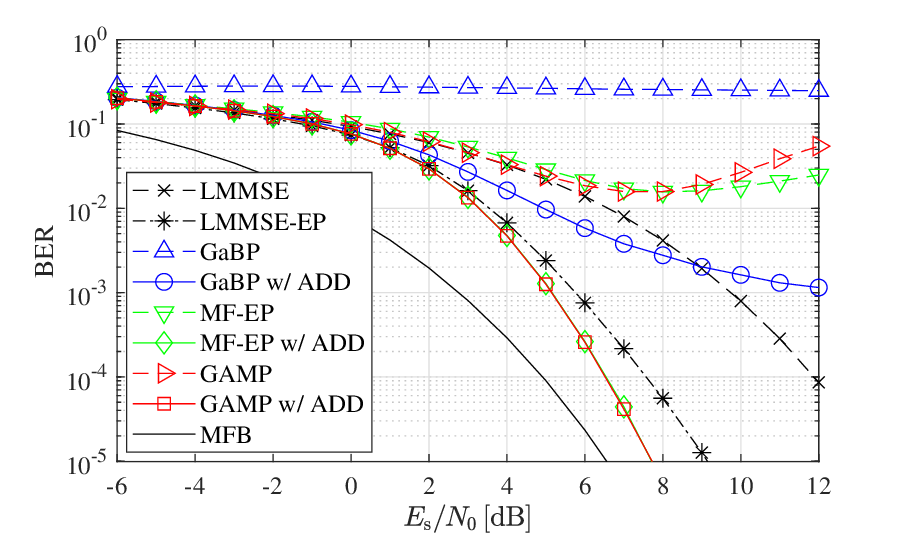}
	\label{fig: ber_16_32_16qam}
	}
	\caption{ \Acs{BER} performance of different detectors in \ac{MU-MIMO} systems with \ac{w.r.t.} $ \Es/N_0 $, where $ (M, N) = (16, 32) $.  }
\label{fig: ber_16_32}
	\vspace{-4mm}
\end{center}
\end{figure*}
%

\begin{figure*}[t]
\begin{center}
\subfloat[$ (M, N) = (24, 32), \, Q = 4, \, \rho = 0.90 $.]{
	\includegraphics[width=0.98\columnwidth,keepaspectratio=true]{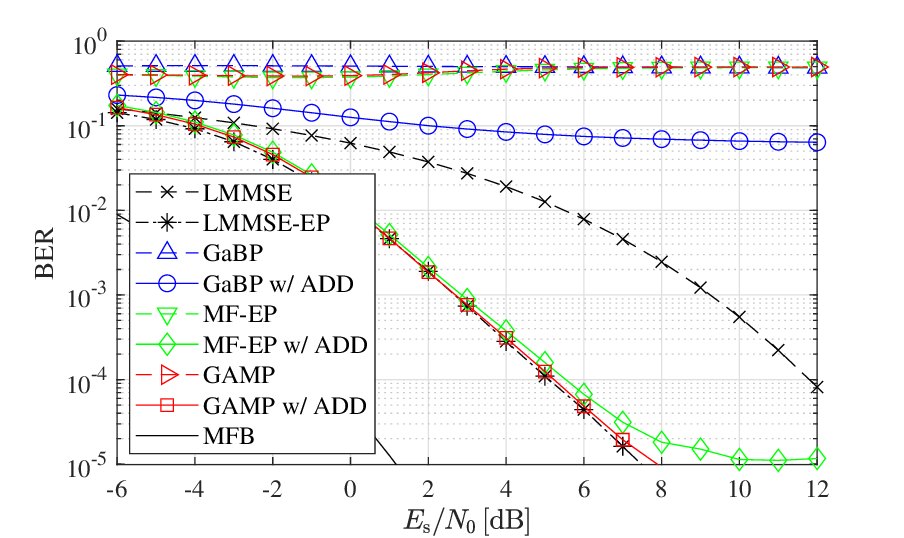}
	\label{fig: ber_24_32_4qam}
	}
 	\hspace{0mm}
	\subfloat[$ (M, N) = (24, 32), \, Q = 16, \, \rho = 0.80 $.]{
	\includegraphics[width=0.98\columnwidth,keepaspectratio=true]{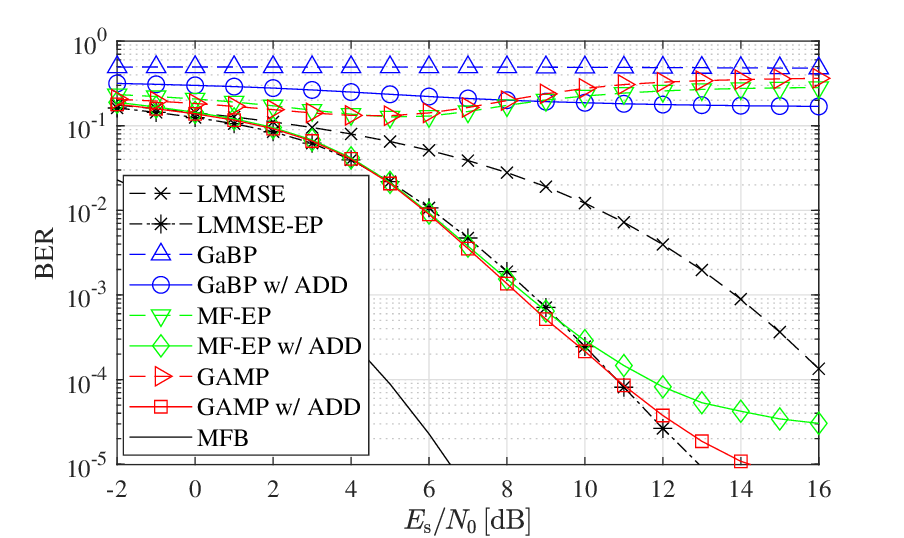}
	\label{fig: ber_24_32_16qam}
	}
	\caption{ \Acs{BER} performance of different detectors in \ac{MU-MIMO} systems with \ac{w.r.t.} $ \Es/N_0 $, where $ (M, N) = (24, 32) $.  }
\label{fig: ber_24_32}
	\vspace{-4mm}
\end{center}
\end{figure*}
%

\begin{figure*}[t]
\begin{center}
\subfloat[$ (M, N) = (64, 64), \, Q = 4, \, \rho = 0.80 $.]{
	\includegraphics[width=0.98\columnwidth,keepaspectratio=true]{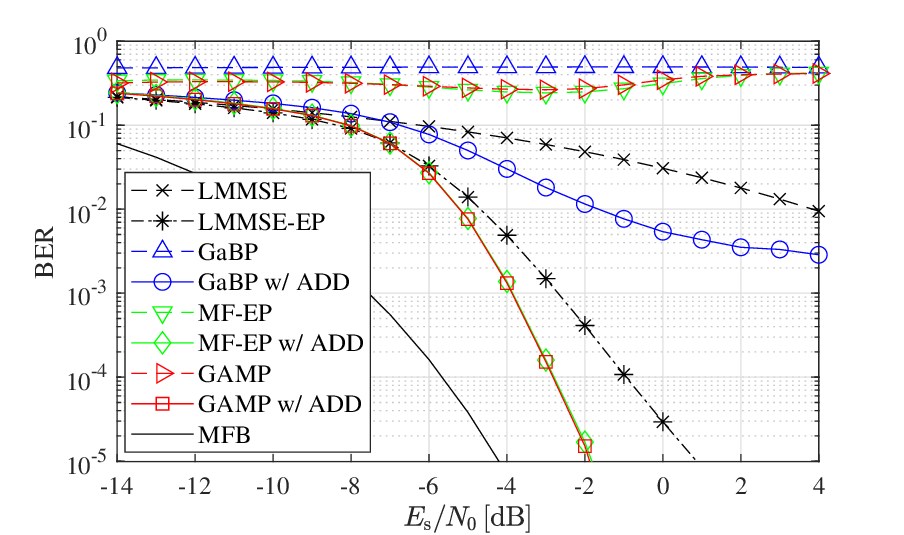}
	\label{fig: ber_64_64_4qam}
	}
 	\hspace{0mm}
	\subfloat[$ (M, N) = (64, 64), \, Q = 16, \, \rho = 0.70 $.]{
	\includegraphics[width=0.98\columnwidth,keepaspectratio=true]{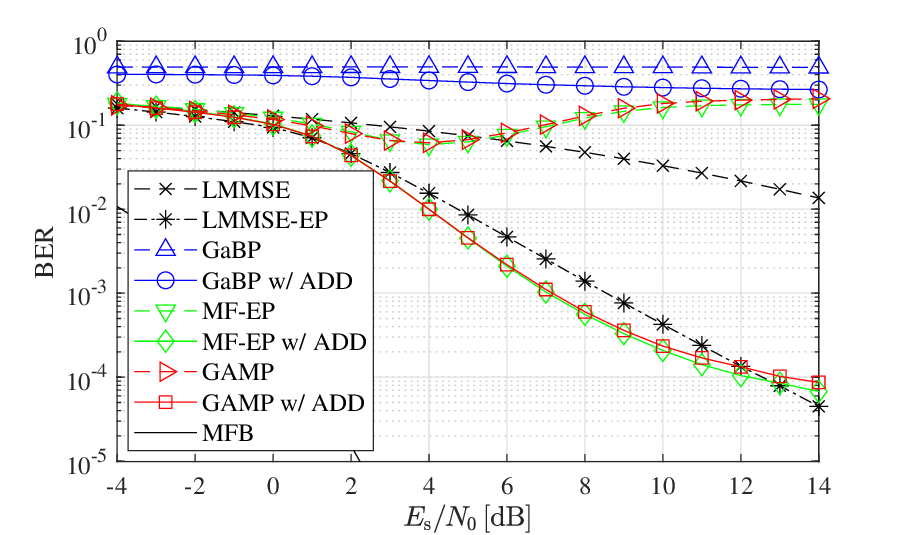}
	\label{fig: ber_64_64_16qam}
	}
	\caption{ \Acs{BER} performance of different detectors in \ac{MU-MIMO} systems with \ac{w.r.t.} $ \Es/N_0 $, where $ (M, N) = (64, 64) $. }
\label{fig: ber_64_64}
	\vspace{-4mm}
\end{center}
\end{figure*}

\subsection{Dynamics of ADD for \ac{QAM} Symbols}

To visualize the dynamics of the denoiser according to the inverse temperature parameter, Fig. \ref{fig: disc_denoiser} shows the dynamics of $\Re\left[\eta_{\beta} \left( y ; 1/\beta \right)\right]$ in \eqref{equ: denoiser_w_asb} with $c^2\cdot\beta \in\left\{0.20,1.0,5.0\right\}$, as a function of $\Re\left[y\right]$, in the cases of $Q=4$ (\textit{i.e.}, $4$-\ac{QAM}) and $Q=16$ ($16$-\ac{QAM}).
The results indicate that the inverse temperature parameter changes the \textit{softness} of the denoiser function, which allows to adjust the convergence rate of \acp{MPA}.
The larger $\beta$ is, the steeper the slope becomes, and the wider the region where the hard-decision symbol is output regardless of the value of input $\Re\left[y\right]$, \textit{i.e.}, \textit{inactive} region.
Based on the above, \cite{Takahashi2019_tcom} suggests that in the early iterations, $\beta$ should be set low to slow down the convergence speed and prevent the error propagation, and in the later iterations, $\beta$ should be set high to promote iterative convergence to the solution along with improving the reliability of beliefs by the \ac{IC}.

The inverse temperature parameter $\beta$ is known to be insensitive to system conditions, and if a simple monotonic function is chosen, the dynamics of $\beta$ are given by a function of the number of iterations as~\cite{Takahashi2019_tcom,Ito2024,Takahashi2023,Takahashi2022,Iimori2021}
\begin{equation}
    \beta^{(t)}
    =
    \frac{d_1}{c^2} \cdot \left( \frac{t}{T} \right)^{d_2},
    \qquad \forall t \in \mathcal{T},
\end{equation}
where $c^2$ is a normalization factor according to the modulation scheme.
In our simulations, $ (d_1, d_2) $ was empirically set to $ (3.0, 2.0) $ via preliminary simulations.

\subsection{ BER Performance }

Our first set of results is given in Figs. \ref{fig: ber_16_32}, \ref{fig: ber_24_32}, and \ref{fig: ber_64_64}, where the \acp{BER} as a function of $ \Es/N_0 $ for the following discrete-valued signal estimation algorithms are compared:
\begin{itemize}
    \item \textit{LMMSE}: Baseline performance of classical signal estimation based on \ac{LMMSE} filtering.
    \item \textit{LMMSE-EP}: A powerful but high-complexity Bayesian optimal \ac{MPA} for unitarily invariant observation, but requires \ac{LMMSE} filtering for each iteration~\cite{Cespedes2014,Takeuchi2020}.
    %
    \item \textit{GaBP (w/ ADD)}:
    Algorithm~\ref{alg: gabp}.
    \item \textit{MF-EP (w/ ADD)}:
    Algorithm~\ref{alg: mfep}.
    \item \textit{GAMP (w/ ADD)}:
    Algorithm~\ref{alg: gamp}.
    \item \textit{MFB}: \Ac{MFB} is the absolute lower bound that the Bayesian \ac{MPA} can ideally achieve if and only if \ac{IC} at the final iteration perfectly works~\cite{Takahashi2022}.
\end{itemize}
For \ac{GaBP}, \ac{MF}-\ac{EP}, and \ac{GAMP}, when using \ac{ADD} (w/ ADD), we assume that \eqref{equ: denoiser_w_asb} is used, and otherwise the conventional Bayesian optimal denoiser in \eqref{equ: discrete_denoiser} is used.
The number of iterations $T$ is set to $64$ 
and the belief damping~\cite{Chockalingam2014} is introduced to the outputs of \ac{LE} with the damping factor of $ 0.50 $ for all the \acp{MPA}.

Figs.~\ref{fig: ber_16_32} and~\ref{fig: ber_24_32} show the \ac{BER} performance with the over-determined system configurations of $ (M, N) = (16, 32) $ and $ (M, N)= (24, 32) $, respectively.
To represent highly correlated measurements, the correlation coefficient is set to $\rho = 0.90$ for $Q=4$ and $\rho = 0.80$ for $Q=16$.
As expected, all the low-complexity \acp{MPA} adopting the conventional Bayesian optimal denoiser in \eqref{equ: discrete_denoiser}, \textit{i.e.},
``GaBP,'' ``MF-EP,'' and ``GAMP,''
equally fail to estimate the desired signals due to the modeling errors in the beliefs caused by the correlation among the elements in $\bm{A}$, and are unable to achieve $ \mathrm{BER} = 10^{-2} $ even in the high $ \Es/N_0 $ region.

On the other hand, by using \ac{ADD} in \eqref{equ: denoiser_w_asb}, ``GaBP w/ ADD'' achieves some improvement in Fig.~\ref{fig: ber_16_32} by avoiding the incorrect hard-decision estimates in the early iterations; however, it still cannot achieve $ \mathrm{BER} = 10^{-4} $ in Fig.~\ref{fig: ber_16_32_4qam} and $\mathrm{BER} = 10^{-3}$ in Fig.~\ref{fig: ber_16_32_16qam}. 
Such improvements are further diluted by a decrease in the compression ratio $\xi = N/M$ (\textit{i.e.}, an increase in spatial load of \ac{MU-MIMO} systems),  as can be seen in Fig.~\ref{fig: ber_24_32}.

In contrast, ``MF-EP w/ ADD'' and ``GAMP w/ ADD'' exhibit significantly improved performance, comparable or even superior to ``LMMSE-EP'' in both Figs.~\ref{fig: ber_16_32} and~\ref{fig: ber_24_32}, without any additional computational cost and/or pre-processing.
In particular, ``GAMP w/ ADD'' shows astonishing robustness even in the case of extremely high correlation ($ \rho = 0.90 $), always achieving $  \mathrm{BER} = 10^{-5}$.
To the best of our knowledge, there is no other method that has successfully implemented an \ac{MPA} with the minimal complexity of order $\mathcal{O}(MN)$ under such conditions.
Notably, the performance degradation from the idealized lower bound ``\ac{MFB}'' at $\mathrm{BER} = 10^{-5}$ is only within approximately $ 2.0 $ dB in Fig.~\ref{fig: ber_16_32_4qam} and even $ 1.5 $ dB in Fig.~\ref{fig: ber_16_32_16qam}, which suggests that \ac{GAMP} has the ability to exchange extrinsic information in a nearly optimal manner even under highly correlated observation conditions by using \ac{ADD}.

In order to investigate the performance in a spatially high loading larger-scale system, in Fig.~\ref{fig: ber_64_64} we increase $M$ and $N$ to $64$ and compare the \ac{BER} performance when $\rho=0.80$ for $Q=4$ and $\rho=0.70$ for $Q = 16$.
Even under these extremely loaded system conditions, ``MF-EP w/ ADD'' and ``GAMP w/ ADD'' still have considerable robustness.
Specifically, in Fig.~\ref{fig: ber_64_64_4qam} they achieve a gain of about $3.0$ dB over ``LMMSE-EP'' at $\mathrm{BER} = 10^{-5}$, and even in the case of Fig.~\ref{fig: ber_64_64_16qam}, which is a substantially overloaded configuration due to higher-order modulation, both ``MF-EP w/ ADD'' and ``GAMP w/ ADD'' can achieve $ \mathrm{BER} = 10^{-4}$, demonstrating the practical applicability of these methods under highly correlated observations.

\subsection{Robustness to Correlation Level}
\begin{figure}[t]
\centering
\includegraphics[width= 1.00\columnwidth,keepaspectratio=true]{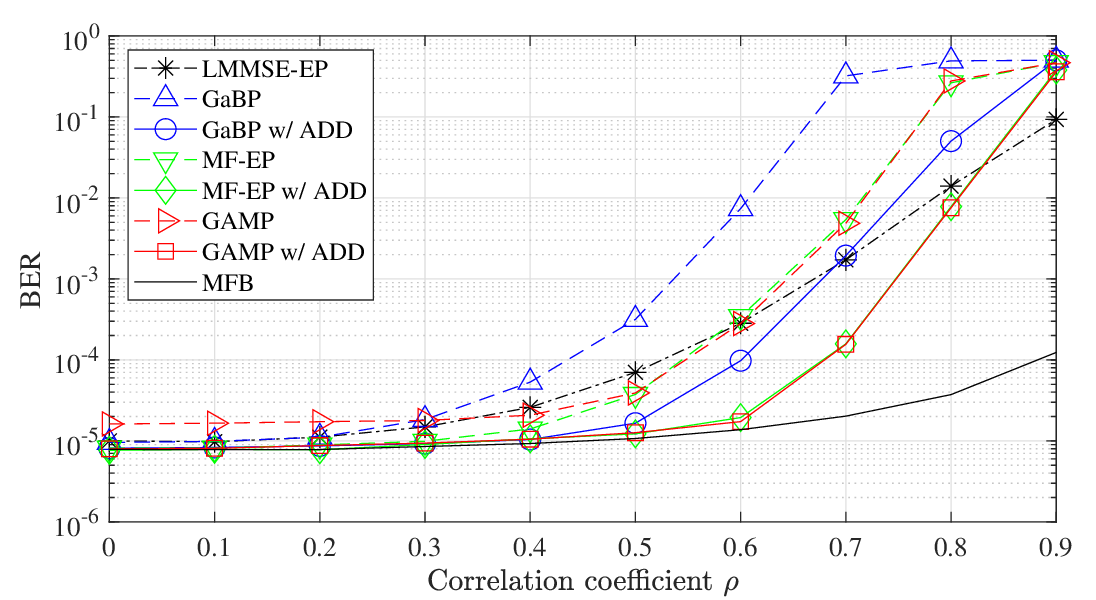}
\caption{BER performance of different detectors in \ac{MU-MIMO} systems \ac{w.r.t.} the correlation coefficient $\rho$ at $\Es/N_0 = -5$ dB, where $ (M, N) = (64, 64), \, Q = 4$, and $T = 64$, respectively.}
\label{fig: ber_rho}
\end{figure}

Next, we comparatively evaluate the robustness of different estimation methods against changes in the correlation level, \textit{i.e.}, $\rho$, in $\bm{A}$.
Fig.~\ref{fig: ber_rho} shows the \ac{BER} performance as a function of the correlation coefficient $\rho$ at $ E_{\mathrm
{s}}/N_0 = -5$ dB under the same system parameter conditions as in Fig.~\ref{fig: ber_64_64_4qam}, \textit{i.e.}, $ (M, N) = (64, 64), \, Q = 4$, and $T = 64$.

It is worth noting that when the correlation coefficient is relatively low (\textit{i.e.}, $\rho \leq 0.50$), all the low-complexity \acp{MPA} outperform ``\ac{LMMSE}-\ac{EP}'' and even asymptotically approach the absolute lower bound ``MFB.''
This result suggests the significant effectiveness of \ac{ADD} in bringing out the full potential of the \ac{MF}-based \acp{MPA} under non-ideal observation conditions.
However, as the value of $\rho$ increases, it can be seen that the \ac{BER} performance of ``GaBP / ADD'' deteriorates rapidly.
In contrast, both ``\ac{MF-EP} w/ ADD'' and ``\ac{GAMP} w/ ADD'' are able to maintain highly accurate estimation even under quite high correlation observations, robustly retaining gain over ``LMMSE-EP'' in the range of $ \rho \leq 0.80$.

\subsection{Convergence Analysis in BER}

\begin{figure}[t]
\centering
\includegraphics[width=1.00\columnwidth,keepaspectratio=true]{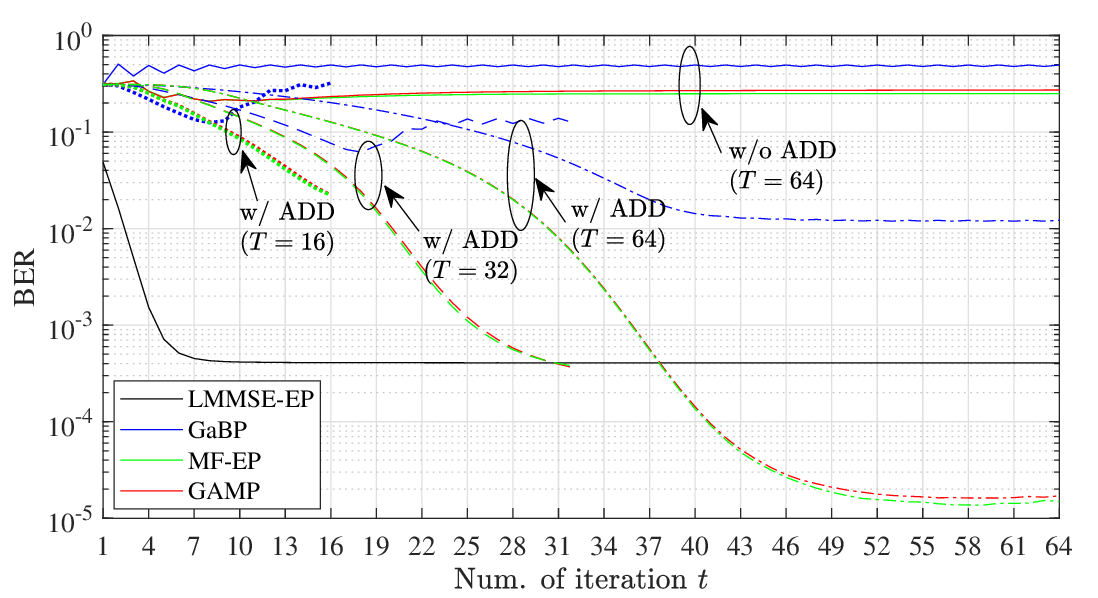}
\caption{BER performance of different detectors in \ac{MU-MIMO} systems \ac{w.r.t.} the iteration number index $t$ for $T = 16, 32,$ and $64$, where $ (M, N) = (64, 64), \, Q = 4, \, \rho = 0.80 $, and $ \Es/N_0 = -2$  dB, respectively.}
\label{fig: ber_ite}
\end{figure}

To evaluate the iterative convergence property of each method, the \ac{BER} performance as a function of the iteration index $t$ at $\Es/N_0 = -2$ dB for different numbers of iterations $T\in\left\{16,32,64\right\}$ is shown in Fig.~\ref{fig: ber_ite}.
The performance of \ac{LMMSE}-\ac{EP}, which iteratively uses the \ac{LMMSE} filtering, is also shown as a reference.
The parameters for the \acp{MPA} are the same as in Figs.~\ref{fig: ber_16_32}-\ref{fig: ber_64_64} and the system configuration is the same as in Fig.~\ref{fig: ber_64_64_4qam}, \textit{i.e.}, $ (M, N) = (64, 64)$, $Q = 4$, and $\rho = 0.80$.

\begin{figure*}[!t]
\centering
\subfloat[$ t=4. $]{\includegraphics[width=0.24\textwidth]{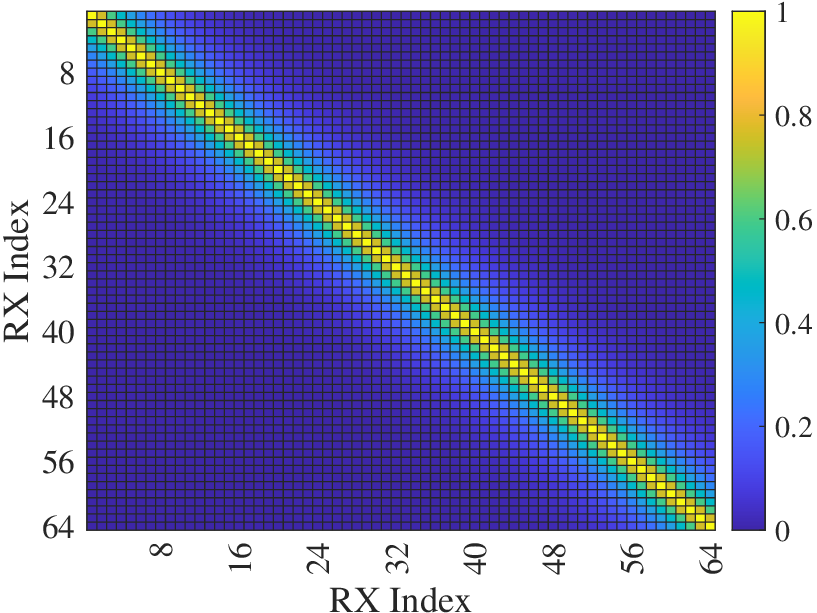}%
\label{fig: cormat_gabp_ite_4}}
\hfil
\subfloat[$ t=20. $]{\includegraphics[width=0.24\textwidth]{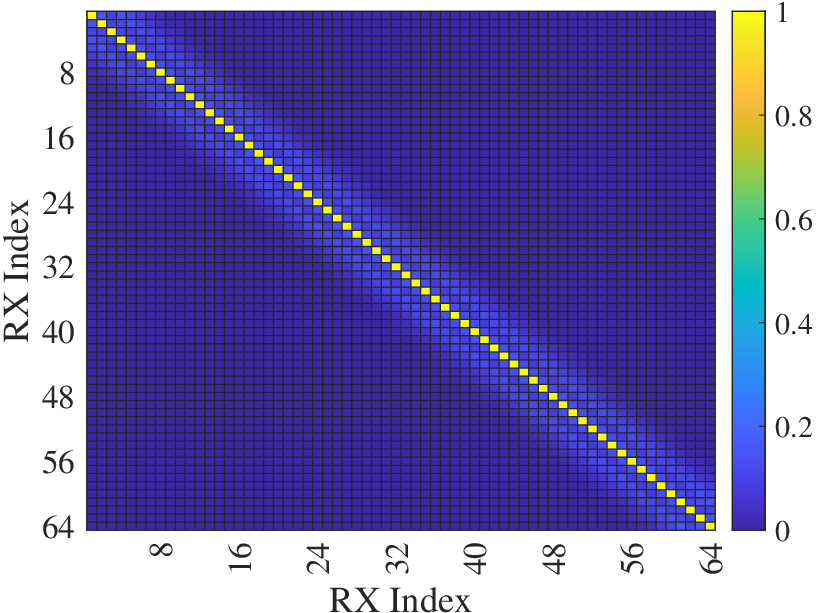}%
\label{fig: cormat_gabp_ite_20}}
\hfil
\subfloat[$ t=40. $]{\includegraphics[width=0.24\textwidth]{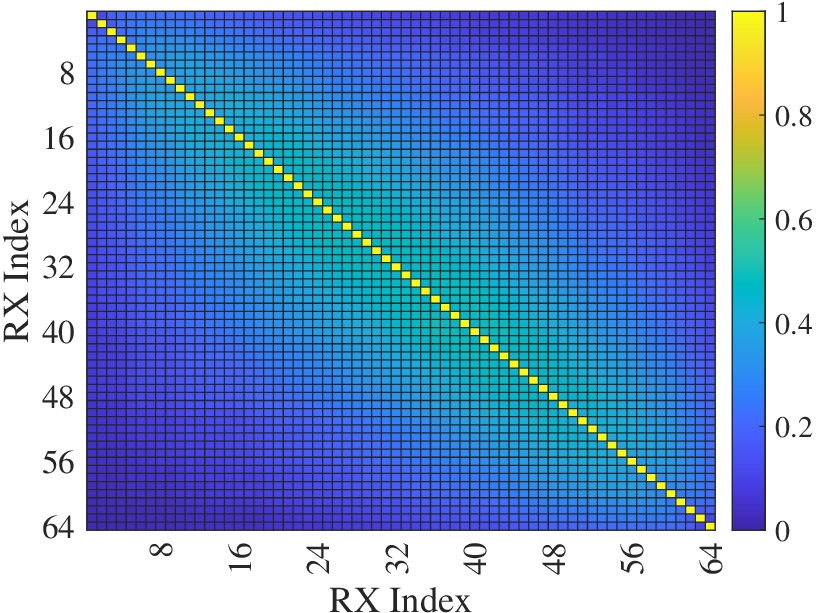}%
\label{fig: cormat_gabp_ite_40}}
\hfil
\subfloat[$ t=60. $]{\includegraphics[width=0.24\textwidth]{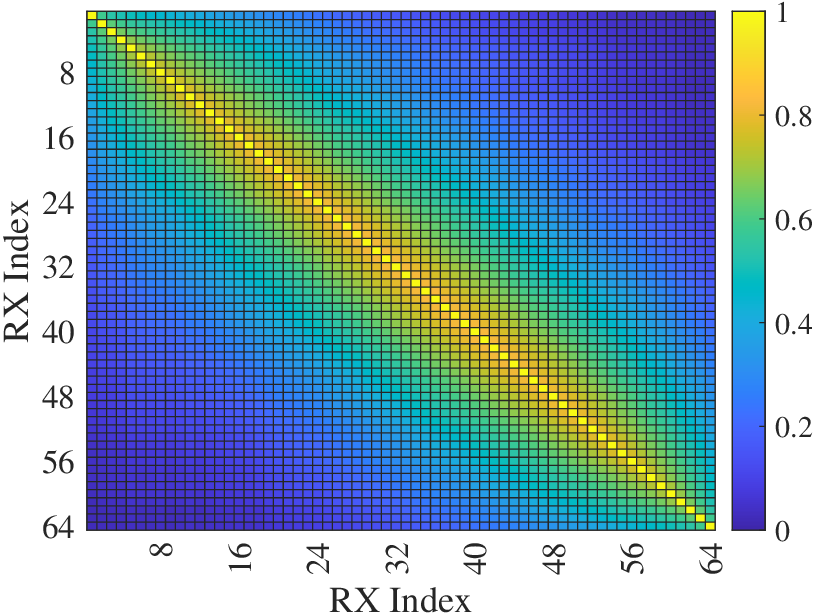}%
\label{fig: cormat_gabp_ite_60}}
\caption{The intensity of the elements in $ \bm{\varGamma}^{(t)} $ for \ac{GaBP} at $ \Es/N_0 = -2$ dB, where $ (M, N) = (64, 64), \rho = 0.80,$ and $Q = 4$.}
\label{fig: gabp_cormat}
\end{figure*}


\begin{figure*}[!t]
\centering
\subfloat[$ t=4. $]{\includegraphics[width=0.24\textwidth]{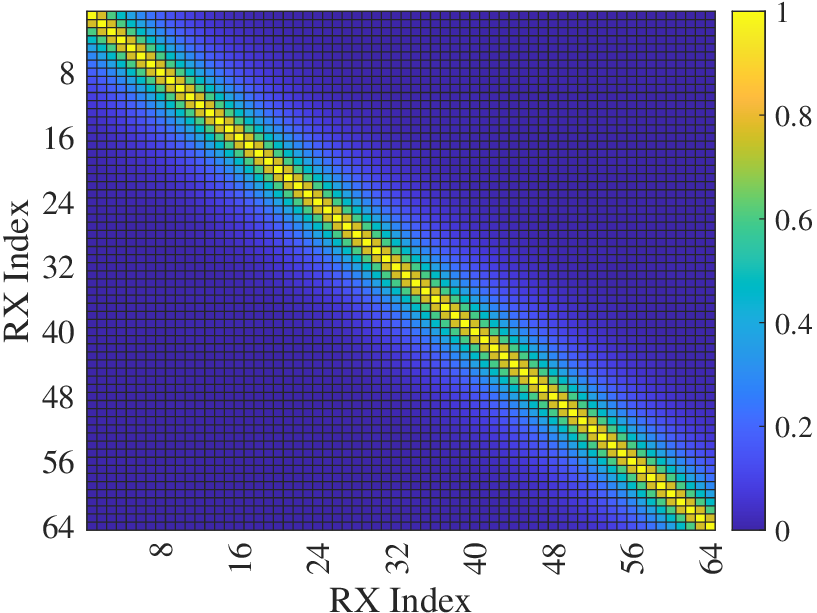}%
\label{fig: cormat_mfep_ite_4}}
\hfil
\subfloat[$ t=20. $]{\includegraphics[width=0.24\textwidth]{FIG/cor_mat/gabp_ite_20_new.eps}%
\label{fig: cormat_mfep_ite_20}}
\hfil
\subfloat[$ t=40. $]{\includegraphics[width=0.24\textwidth]{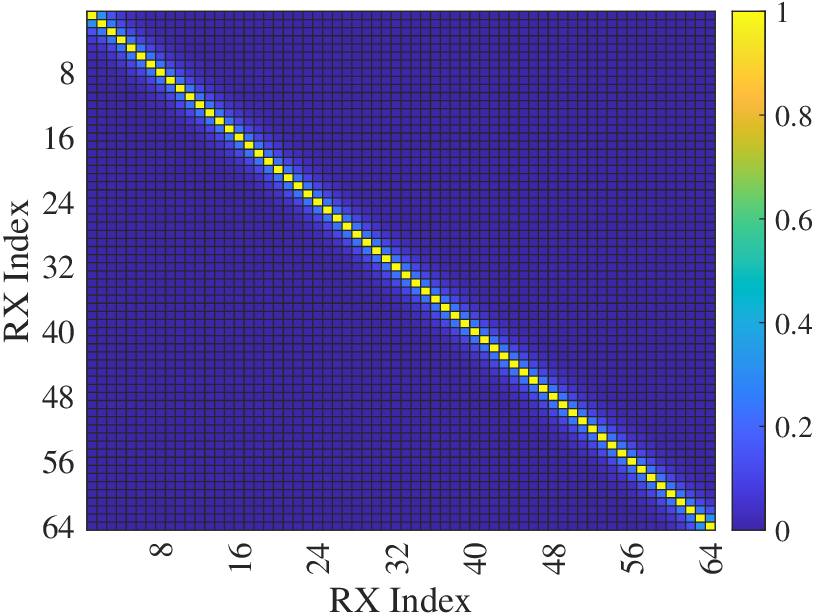}%
\label{fig: cormat_mfep_ite_40}}
\hfil
\subfloat[$ t=60. $]{\includegraphics[width=0.24\textwidth]{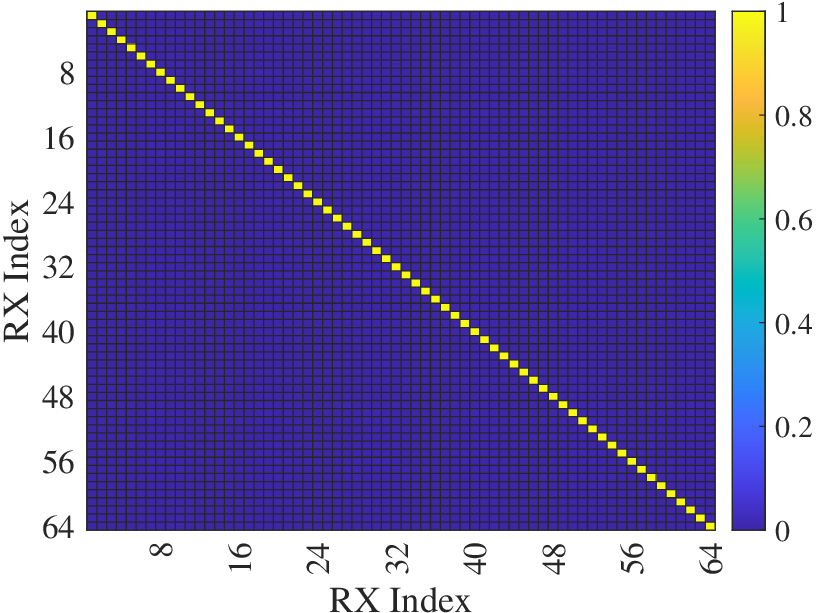}%
\label{fig: cormat_mfep_ite_60}}
\caption{The intensity of the elements in $ \bm{\varGamma}^{(t)} $ for \ac{MF-EP} at $ \Es/N_0 = -2$ dB, where $ (M, N) = (64, 64), \rho = 0.80,$ and $Q = 4$.}
\label{fig: mfep_cormat}
\end{figure*}


\begin{figure*}[!t]
\centering
\subfloat[$ t=4. $]{\includegraphics[width=0.24\textwidth]{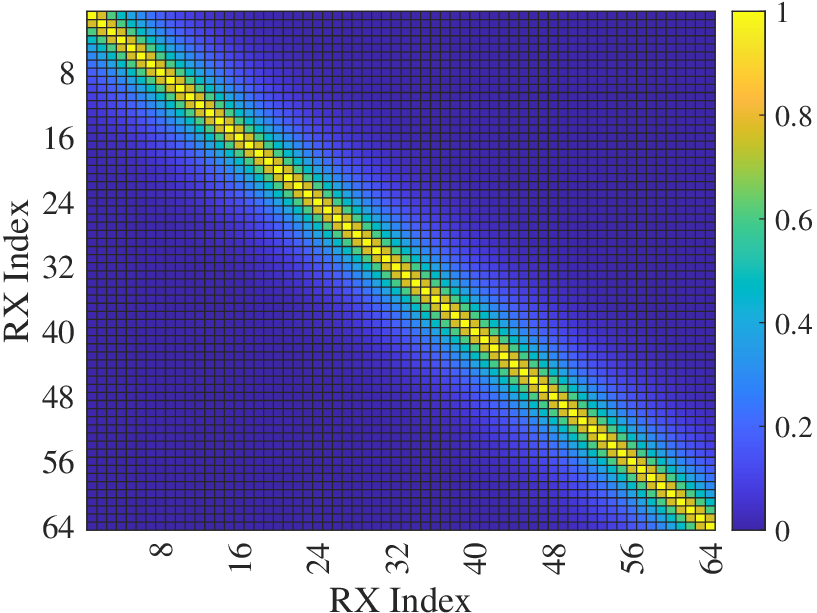}%
\label{fig: cormat_gamp_ite_4}}
\hfil
\subfloat[$ t=20. $]{\includegraphics[width=0.24\textwidth]{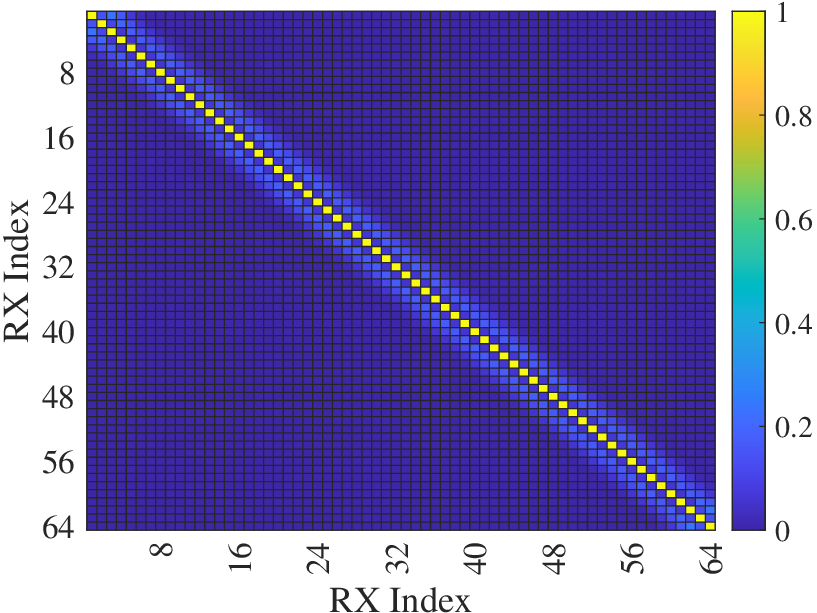}%
\label{fig: cormat_gamp_ite_20}}
\hfil
\subfloat[$ t=40. $]{\includegraphics[width=0.24\textwidth]{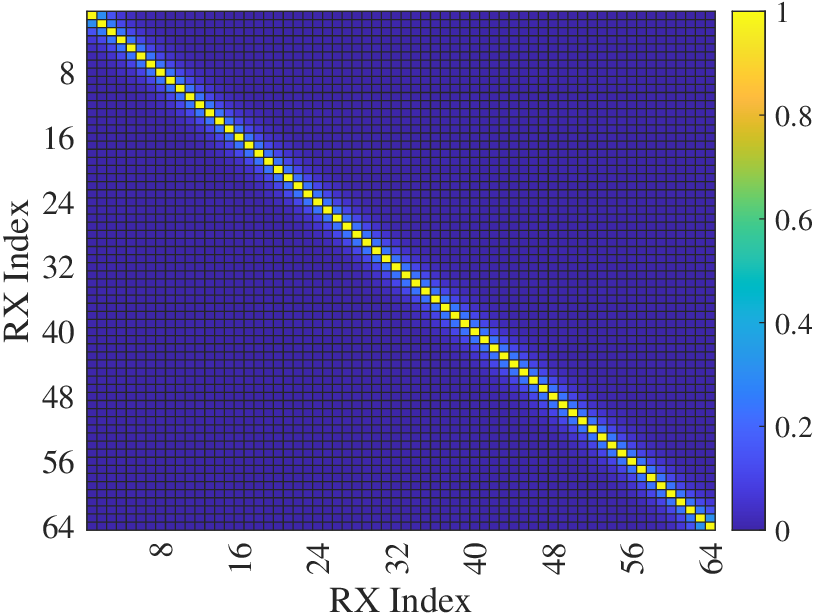}%
\label{fig: cormat_gamp_ite_40}}
\hfil
\subfloat[$ t=60. $]{\includegraphics[width=0.24\textwidth]{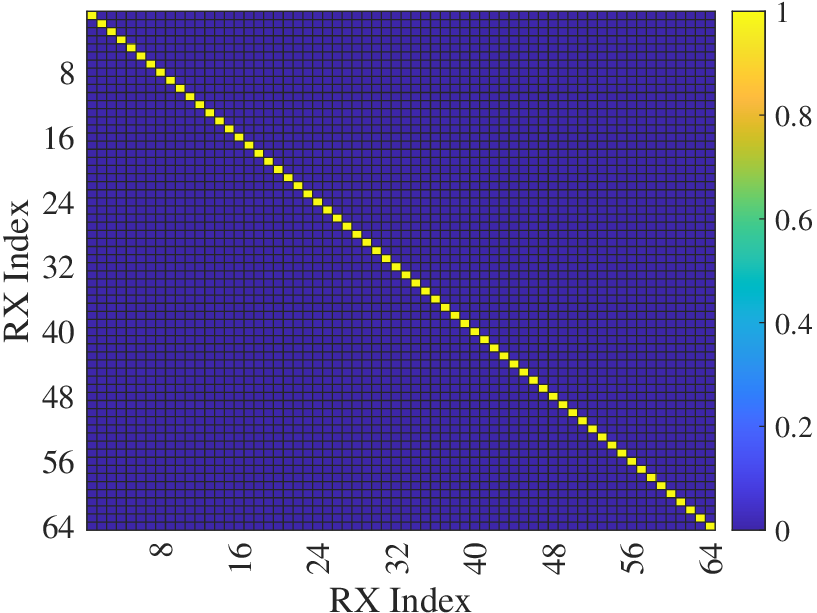}%
\label{fig: cormat_gamp_ite_60}}
\caption{The intensity of the elements in $ \bm{\varGamma}^{(t)} $ for \ac{GAMP} at $ \Es/N_0 = -2$ dB, where $ (M, N) = (64, 64), \rho = 0.80,$ and $Q = 4$.}
\label{fig: gamp_cormat}
\end{figure*}

In contrast to the case of ``\ac{GaBP}'', the \ac{BER} decreases monotonically with the \ac{BER} in the early iterations of ``\ac{GaBP} w/ ADD,'' which means that the use of \ac{ADD} can suppress the error propagation to some extent.
However, in the case of $T=64$, the decrease in the \ac{BER} stalls as the number of iterations exceeds some level, and in the case of $T=16$ and $32$, a V-shaped curve appears, indicating an increase in \ac{BER}.
This suggests that the estimation process converges to undesirable local minima due to the harmful effects of error propagation.
In contrast, the \acp{BER} of ``\ac{MF-EP} w/ ADD'' and ``\ac{GAMP} w/ ADD'' are seen to decrease steadily throughout the entire iterative process as the inverse temperature parameter $\beta^{(t)}$ increases.
These results suggest that, in the case of \ac{MF}-\ac{EP} and \ac{GAMP}, the exchange of extrinsic information is successful across iterations with the aid of \ac{ADD}, but this is not the case for \ac{GaBP}.

To gain an insight on this interesting observation, in the following subsection, we will analyze the stochastic behavior of beliefs propagating in each of \ac{GaBP}, \ac{MF-EP}, and \ac{GAMP}.

\subsection{Iterative Behavior Analysis}


\begin{figure*}[t]
\begin{eqnarray}
\label{equ: def_e_n}
e_n^{(t)} \triangleq
\begin{cases}
\displaystyle
y_n - \sum_{m=1}^{M} a_{n, m} \check{x}_{n, m}^{(t)}
=
\sum_{m=1}^{M} a_{n, m}
\left( x_m -  \check{x}_{n, m}^{(t)} \right) + w_n
,  & \textrm{for \ac{GaBP} and \ac{MF-EP},}\\
\displaystyle
y_n - \sum_{m=1}^{M} a_{n, m} \check{x}_{m}^{(t)} + \gamma_{n}^{(t)} s_{n}^{(t-1)}
=
\sum_{m=1}^{M} a_{n, m} \left( x_m -\check{x}_{m}^{(t)} \right) + \gamma_n^{(t)} s_n^{(t-1)}
+
w_n
, & \textrm{for \ac{GAMP}.}\\
\end{cases}
\end{eqnarray}
\hrule
\vspace{-2mm}
\end{figure*}

As we have seen in Sections~\ref{chap: gabp_and_mfep} and \ref{chap: derivation_of_gamp}, the crucial difference among \ac{GaBP}, \ac{MF-EP}, and \ac{GAMP} lies in the mechanism for suppressing the self-noise feedback, which underpins the fundamental operating principle of Bayesian optimal \acp{MPA}.
Specifically, \ac{GaBP} performs the self-noise suppression by excluding the self-feedback component in the belief combining process of (A\ref{alg: gabp}-5) and (A\ref{alg: gabp}-6), \ac{MF-EP} performs the self-noise suppression in the the soft replica dimension after denoising by \ac{MM} of (A\ref{alg: mfep}-9) and (A\ref{alg: mfep}-10), and \ac{GAMP} performs the self-noise suppression in \ac{LE} in the next iterative process by the Onsager correction term of (A\ref{alg: gamp}-5).
It is worth noting here that the \textit{timing} of the self-noise suppression is different in the iterative process.
More precisely, in \ac{GaBP}, self-noise suppression is performed before denoising, while in \ac{MF-EP} and \ac{GAMP}, it is performed after denoising.
In fact, this difference in the relative position of the processing within the algorithm offers the completely different level of robustness against highly correlated observations, as described below.

In \ac{GaBP}, where the self-noise suppression mechanism is performed immediately before the denoiser, it is quite difficult to appropriately propagate extrinsic information in the presence of the modeling errors of beliefs.
This is because when the denoiser begins to operate in the \textit{inactive} region shown in Fig. \ref{fig: disc_denoiser} due to noise enhancement arising from the severe correlations in $\bm{A}$, the setting of a large inverse temperature parameter $\beta$, or both, one loses the \textit{effect of excluding the self-feedback component in extrinsic belief combining} that should be propagated to the next iteration.
This causes almost the same soft replicas to be fed back to all the \acp{FN} in the next iteration (\textit{i.e.}, $\check{x}_{1, m} \simeq \check{x}_{2, m} \simeq \cdots \simeq \check{x}_{N, m} $), which violates the fundamental operating principle of \acp{MPA} based on the extrinsic information exchange, resulting in significant performance degradation.
In contrast, in \ac{MF-EP} and \ac{GAMP}, the self-noise feedback is removed after the denoiser; therefore, even if the denoiser operates somewhat irrationally in the inactive region, the extrinsic information can still be exchanged across iterations, allowing accurate discrete-valued signal estimation using \ac{ADD}, even for highly correlated measurements.
%


\begin{figure*}[!t]
\centering
\subfloat[GaBP w/ ADD]{\includegraphics[width=0.33\textwidth]{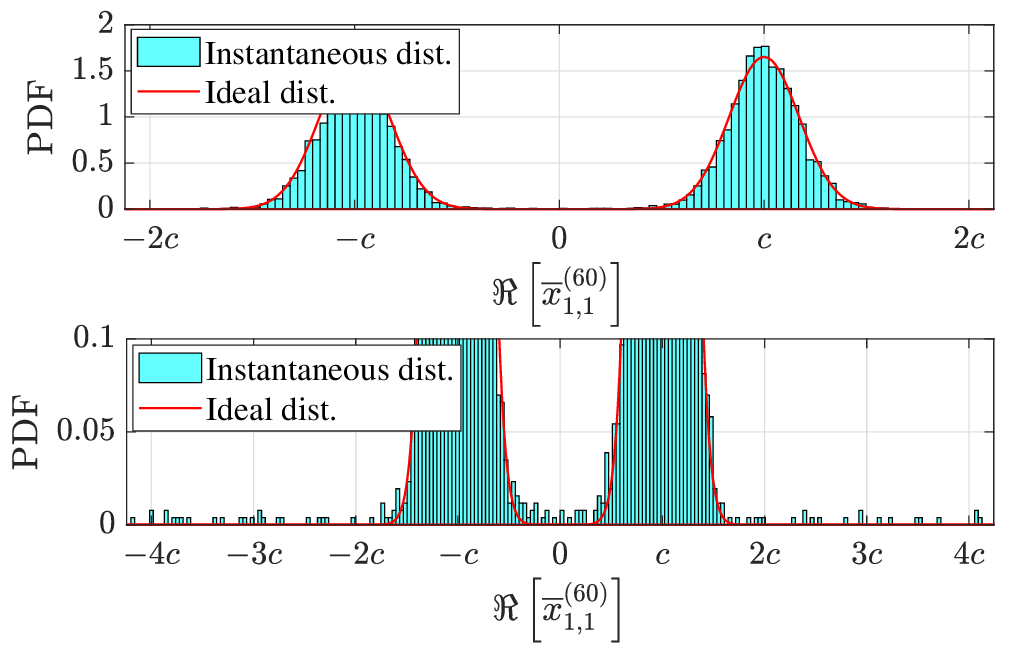}%
\label{fig: ber_dist_gabp}}
\hfil
\subfloat[MF-EP w/ ADD]{\includegraphics[width=0.33\textwidth]{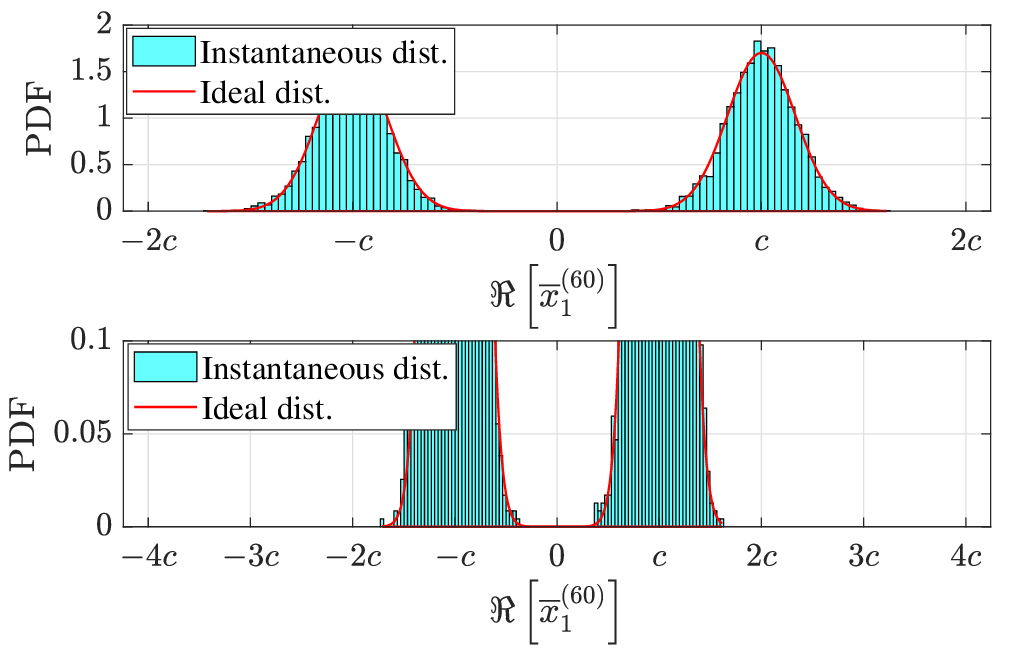}%
\label{fig: ber_dist_mfep}}
\hfil
\subfloat[GAMP w/ ADD]{\includegraphics[width=0.33\textwidth]{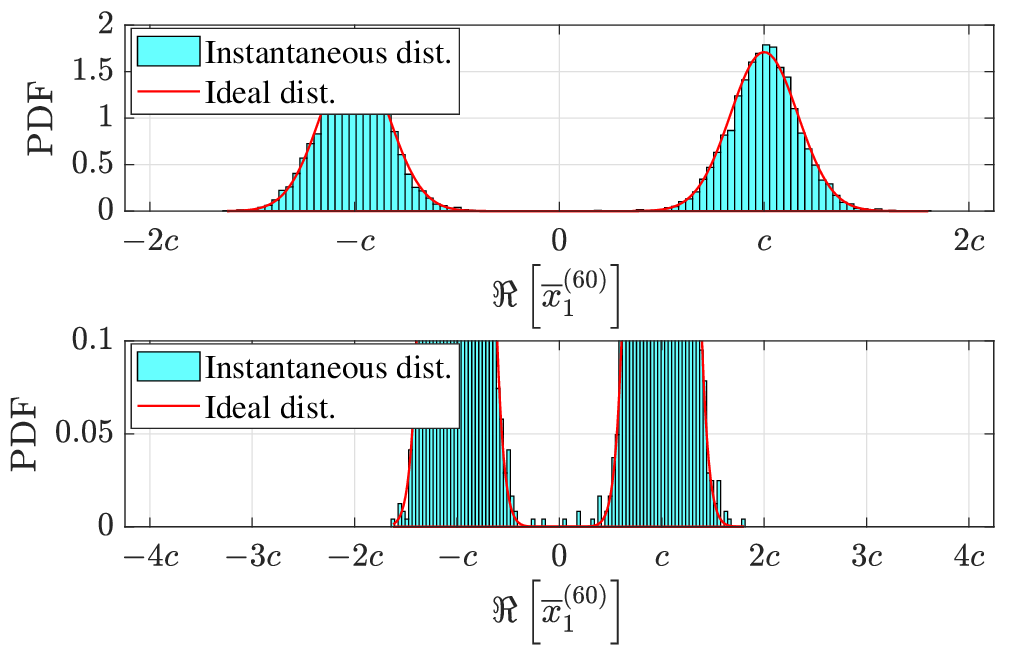}%
\label{fig: ber_dist_gamp}}
\caption{The histogram of \ac{LE} output corresponding to Fig.~\ref{fig: cormat_gabp_ite_60}, \ref{fig: cormat_mfep_ite_60}, and \ref{fig: cormat_gamp_ite_60}. }
\label{fig: b_dist}
\vspace{-4mm}
\end{figure*}

To verify the above analytical conjecture, we numerically analyze the stochastic behavior of beliefs in each of \ac{GaBP}, \ac{MF-EP}, and \ac{GAMP}, under highly correlated measurements.
These three algorithms all operate under the assumption that the effective noise after \ac{IC}, as defined in \eqref{equ: def_e_n}, is uncorrelated among observations as a result of appropriate self-noise suppression; hence, to verify this, we will focus on the correlation matrix $\bm{\varGamma}^{(t)} \in \mathbb{C}^{N \times N}$ whose $(i, j)$-th entry is defined as
\begin{equation}
\left[\bm{\varGamma}^{(t)} \right]_{i, j}
=
\frac{ \mathbb{E}_{ \bm{\mathsf{x}} , \bm{\mathsf{A}} , \bm{\mathsf{w}}  } \left[ \bigl(e_i^{(t)}\bigr)^* e_j^{(t)} \right]  }{
\sqrt{
\mathbb{E}_{ \bm{\mathsf{x}} , \bm{\mathsf{A}} , \bm{\mathsf{w}}  } \left[ \bigl|e_i^{(t)} \bigr|^2 \right]
}
\sqrt{
\mathbb{E}_{ \bm{\mathsf{x}} , \bm{\mathsf{A}} , \bm{\mathsf{w}}  } \left[ \bigl|e_j^{(t)} \bigr|^2 \right]
}
}.
\end{equation}
In other words, the closer $\bm{\varGamma}^{(t)}$ is to a diagonal matrix, the more successful the appropriate extrinsic information exchange is, and ideally it will be an identity matrix.

Figs.~\ref{fig: gabp_cormat},~\ref{fig: mfep_cormat}, and~\ref{fig: gamp_cormat} show the intensity of each element in $ \bm{\varGamma}^{(t)} $ under the same parameters as in Fig.~\ref{fig: ber_ite}, \textit{i.e.}, $ (M, N) = (64, 64), \, Q = 4, \, \rho = 0.80$, $T=64$, and $ \Es/N_0 = -2$\, dB.
In the early iterations (\textit{e.g.}, $t=4$) of Figs.~\ref{fig: cormat_gabp_ite_4}, \ref{fig: cormat_mfep_ite_4}, and \ref{fig: cormat_gamp_ite_4},
$ \bm{\varGamma}^{(4)} $ shows a structure in which the intensity of elements decays exponentially as they move away from the main diagonal elements in all \acp{MPA}, which is clearly due to the correlation matrix $ \bm{A}$ given in \eqref{equ: R_rx_exponentially_decaying}.
As the iterative process progresses, in Figs.~\ref{fig: cormat_gabp_ite_20},~\ref{fig: cormat_mfep_ite_20}, and~\ref{fig: cormat_gamp_ite_20} for $t = 20$,
the extrinsic information exchange is enabled by \ac{ADD}, and the statistical independence among beliefs improves in most cases except in the vicinity of the diagonal elements of $\bm{\varGamma}^{(20)}$.
However, in the later iterations (\textit{i.e.}, $t = 40$ and $60$), when the inverse temperature parameter $\beta^{(t)}$ needs to be set to a large value to promote convergence, \ac{GaBP} is unable to exchange extrinsic information again due to the modeling errors remaining in the belief, and the off-diagonal elements of $\bm{\varGamma}^{(t)}$ increase as shown in Figs.~\ref{fig: cormat_gabp_ite_40} and~\ref{fig: cormat_gabp_ite_60}.
This is the cause of the V-shaped curve and the stagnation of the \ac{BER} reduction in Fig. \ref{fig: ber_ite}.
In contrast, \ac{MF-EP} and \ac{GAMP} can maintain the extrinsic information exchange, as can be seen from almost the identity matrix $\bm{\varGamma}^{(60)}$ in 
Figs.~\ref{fig: cormat_mfep_ite_60} and \ref{fig: cormat_gamp_ite_60}.
This indicates that the self-noise suppression mechanism based on \ac{MM}/Onsager correction performed after denoising can be robust against the belief modeling errors.

To provide a more intuitive understanding of the detrimental effects of harmful self-feedback in \ac{GaBP},
Fig.~\ref{fig: b_dist} shows the normalized histograms of the beliefs (\ac{LE} outputs) corresponding to Figs.~\ref{fig: cormat_gabp_ite_60},~\ref{fig: cormat_mfep_ite_60}, and~\ref{fig: cormat_gamp_ite_60}.
In addition, the ideal estimated Gaussian distribution assumed by the denoiser is also depicted in the red lines.
At first glance, the normalized histograms (Empirical dist.) of all the \acp{MPA} in the upper part of Fig.~\ref{fig: b_dist} appear to be consistent with the ideal Gaussian distribution (Ideal dist.). However, if we focus on the tails of the distributions shown in the lower part, the presence of \textit{belief outliers} due to modeling errors becomes apparent.
Specifically, in Fig.~\ref{fig: ber_dist_gabp}, the histogram has an abnormally heavy tail, which is the direct cause of the performance degradation observed in \ac{GaBP}.
In contrast, almost no such outliers are observed in Figs.~\ref{fig: ber_dist_mfep} and~\ref{fig: ber_dist_gamp} of \ac{MF-EP} and \ac{GAMP}, which means that, with appropriate extrinsic information exchange, the stochastic behavior of beliefs follows the ideal Gaussian distribution based on the \ac{AWGN} observation model.

\section{Conclusion}

In this paper, we have proposed an advanced and effective strategy that enables accurate discrete-valued signal estimation via the Bayesian optimal \acp{MPA} with the minimal complexity under highly correlated observations.
To elucidate the structural differences and interrelationships among \ac{GaBP}, \ac{MF-EP}, and \ac{GAMP}, we have shown the systematic derivation of \ac{GAMP} from \ac{SPA} via \ac{GaBP} in the complex-domain, and clarified the differences in their self-noise suppression mechanisms.
Numerical results have demonstrated that these three algorithms with \ac{ADD} exhibit significantly different estimation accuracies under the highly correlated measurements.
Finally, we have conjectured the cause of this phenomenon in terms of their different self-noise suppression mechanisms, which has been described theoretically through the analysis of the statistical correlation structure among the beliefs.
The above findings pave the way for the practical applicability of the low-complexity \ac{MPA} even under correlated measurements.

\appendices

\section{Proof of Lemma \ref{lem: complex_feedback_0}}
\label{chap: proof_of_lemma}

For later convenience, we first prove $ \mathbb{E}_{\mathsf{a}_{n, m}} \left[a_{n, m}^2 \right] = 0 $ for every entry of $\bm{A}$ as follows:
when $ a_{n, m} $ is circularly symmetric,
$ \mathbb{E}_{\mathsf{a}_{n, m}} \left[a_{n, m}^2 \right] = \mathbb{E}_{\mathsf{a}_{n, m}} \left[ a_{n, m}^2 e^{\mathrm{j} 2\theta} \right] $, \textit{i.e.}, 
$ \left( 1 - e^{\mathrm{j} 2 \theta } \right) \mathbb{E}_{\mathsf{a}_{n, m}} \left[a_{n, m}^2 \right] = 0 $ holds for all $ \theta \in [0, 2\pi) $, which necessitates $ \mathbb{E}_{\mathsf{a}_{n, m}} \left[a_{n, m}^2 \right] = 0 $ since 
$ \left( 1 - e^{\mathrm{j} 2 \theta } \right) $ is not always 0.
In addition, based on the assumption that $ \check{r}_1, \check{r}_2 \ldots , \check{r}_M  $ are all bounded, there exists $ R < \infty $ such that $ \left| \check{r}_{m} \right| < R $ for all $ 
m \in \left\{1, 2, \cdots , M \right\} $.
Therefore, we have
\begin{eqnarray}
&&
\!\!\!\! 
 \mathbb{E}_{\bm{\mathsf{A}}}
 \left[ \left| 
 \sum_{m=1}^{M}  a_{n, m}^2 \check{r}_m
 \right|^2 
 \right]
 <
 R^2
 \cdot
 \mathbb{E}_{\bm{\mathsf{A}}}
 \left[ \left| 
 \sum_{m=1}^{M}  a_{n, m}^2
 \right|^2 
 \right]
 \nonumber
 \\
 \!\!\!\!&=&\!\!\!\!
    R^2 \cdot
 \mathbb{E}_{\bm{\mathsf{A}}}
 \left[ 
 \sum_{m=1}^{M}
 \sum_{m'=1}^{M}
 a_{n, m'}^2
 (a_{n, m}^*)^2
 \right]
\nonumber
\\
\!\!\!\!&=&\!\!\!\!
 R^2 \cdot
 \sum_{m=1}^{M}
 \overbrace{
 \mathbb{E}_{\mathsf{a}_{n,m}}
 \left[ 
 |a_{n, m}|^4
 \right]
 }^{= \mathcal{O}(N^{-2})}
 \nonumber
 \\
 &&
 +
 \,
 R^2 \cdot
\sum_{m=1}^{M}
 \sum_{m' \neq m}^{M}
 \underbrace{
 \mathbb{E}_{\mathsf{a}_{n, m'}}
 \left[ 
 a_{n, m'}^2
 \right]
 }_{=0}
 \underbrace{
 \left(
 \mathbb{E}_{\mathsf{a}_{n, m}}
 \left[
 a_{n, m}^2
 \right]
 \right)^*
 }_{=0}
 \nonumber
 \\
 \!\!\!\!&=&\!\!\!\!
 \mathcal{O}\left(N^{-1}\right),
\end{eqnarray}
where the last equality holds since $N, M \to \infty$ with $\xi = N/M$ fixed in the large-system limit.



\bibliographystyle{REF/IEEEtran}
\bibliography{REF/ref}

\end{document}